\documentclass[reprint,superscriptaddress,nofootinbib,nobibnotes,amsmath,amssymb,aps,prd]{revtex4-2}

\usepackage[colorlinks=true,linkcolor=blue,citecolor=blue,]{hyperref}
\usepackage{amsmath,amssymb,amsthm,amsfonts,mathrsfs,bm}
\usepackage{cprotect}
\usepackage{graphicx,tabularx}
\graphicspath{{fig/}}
\usepackage{rotating}
\usepackage{float}
\usepackage{multirow}
\usepackage{cancel}
\usepackage{comment}
\usepackage{physics}
\usepackage{array}
\usepackage{enumitem}
\newcolumntype{P}[1]{>{\centering\arraybackslash}p{#1}}

\usepackage{xcolor}
\usepackage{dcolumn}% Align table columns on decimal point

\usepackage{iftex}
\ifpdftex% Commands and packages specific to pdfLaTeX
    \usepackage{CJKutf8}
    \newcommand{\Ckai}[1]{\begin{CJK*}{UTF8}{gkai}#1\end{CJK*}}
\fi
\ifxetex% Commands and packages specific to XeLaTeX
    \usepackage{xeCJK}
    \newcommand{\Ckai}[1]{\textit{#1}}
\fi

\newcommand{\eqcontrimark}{\star}
\newcommand{\eqcontri}{$^{\color{blue}\eqcontrimark}$}

\newcommand{\BH}{\mathrm{BH}}

\begin{document}
\title{Boson Stars Hosting Black Holes}

\author{Amitayus Banik\eqcontri}
\email{abanik@cbnu.ac.kr}
\affiliation{Department of Physics, Chungbuk National University, Cheongju, Chungbuk 28644, Korea}
\affiliation{Research Institute for Nanoscale Science and Technology, Chungbuk National University, Cheongju, Chungbuk 28644, Korea}

\author{Jeong Han Kim\eqcontri}
\email{jeonghan.kim@cbu.ac.kr}
\affiliation{Department of Physics, Chungbuk National University, Cheongju, Chungbuk 28644, Korea}

\author{Xing-Yu Yang (\Ckai{杨星宇})}
\email[Corresponding author:~]{xingyuyang@kias.re.kr}
\affiliation{Quantum Universe Center (QUC), Korea Institute for Advanced Study, Seoul 02455, Republic of Korea}

\def\thefootnote{$\eqcontrimark$}
\footnotetext{These authors contributed equally to this work.}
\def\thefootnote{\arabic{footnote}}
\setcounter{footnote}{0}

\begin{abstract}
    We study a self-gravitating ultralight dark matter condensate (a boson star) hosting a central black hole, in the nonrelativistic limit, which we refer to as a boson star black hole (BS-BH) system. We numerically solve the equations of hydrostatic equilibrium, consistently incorporating the gravitational potential of the black hole, to obtain all possible configurations of this BS-BH system for different boson star masses, interaction types, and black hole masses. We also propose an analytic expression for the density profile and compare it with the numerical results, finding good agreement for attractive interactions and for a finite range of mass ratios between the black hole and boson star. Finally, considering the inspiral of this BS-BH system with a second, smaller black hole, we study the dephasing of gravitational waves due to the presence of the dark matter environment. A Fisher matrix analysis reveals the regions of parameter space of the dark matter mass and self-coupling that future gravitational wave observatories such as LISA can probe.
\end{abstract}

\maketitle

\section{Introduction}
\label{sec:intro}

Whether the scalar particle stands alone or there are other scalar degrees of freedom still remains an open question and continues to draw considerable interest among theorists.
In fact, many beyond the Standard Model scenarios generically predict additional scalar particles, such as the QCD axion, which was proposed to address the strong CP problem~\cite{Peccei:1977hh, Peccei:1977ur, Weinberg:1977ma, Wilczek:1977pj}, and axionlike particles motivated from string theory~\cite{Witten:1984dg, Conlon:2006tq, Svrcek:2006yi, Arvanitaki:2009fg, Choi:2009jt, Acharya:2010zx, Marsh:2011gr, Higaki:2011me, Cicoli:2012sz, Halverson:2017deq, Demirtas:2018akl}. These new scalar particles are compelling from both theory and experimental viewpoints, because they can also provide candidates for dark matter (DM), a missing ingredient in the Standard Model.

In this work, we consider self-interacting ultralight dark matter (ULDM, cf. Refs.~\cite{Ferreira:2020fam,Hui:2021tkt, Chavanis:2025qcg} for exhaustive reviews), modeled as a real scalar field with mass $m$ and quartic self-coupling $\lambda$. A defining feature of such bosonic particles is their ability to form stable and self-gravitating condensates, which are referred to as solitons or boson stars.
This stable configuration is formed from the balance between the quantum-mechanical pressure, the pressure due to self-interactions and the attractive gravitational force of the system. Depending on $m$, $\lambda$ and the total mass of the boson star, its characteristic size can span several orders of magnitude, from $\mathcal{O}(1000~{\rm pc})$ to $\mathcal{O}(100~{\rm km})$~\cite{Chavanis:2025qcg}, thereby providing leverage across multiple astrophysical probes.

As a concrete astrophysical probe, we examine systems in which a black hole sits at the center of the boson star core, referred to hereafter as boson star black hole (BS-BH) systems. Such configurations can arise, for example, when primordial black holes seed the growth of ULDM miniclusters that relax into boson stars in their vicinity~\cite{Hertzberg:2020hsz,Yin:2024xov,Palomares-Chavez:2024liz}.
Depending on the black hole mass scale, these BS-BH systems admit distinct observational handles.
For supermassive black holes (SMBHs)~\cite{Kormendy:1995er,Ferrarese:2004qr} with masses of $ \sim 10^6 - 10^{10} M_{\odot}$, an overdense scalar environment can modify stellar orbits, and can shift the angular size of the SMBH shadow.
Moreover, accretion onto the SMBH can deplete the boson star configuration.
Several studies~\cite{Bar:2019pnz,Davies:2019wgi,Chakrabarti:2022owq, Gan:2023swl} have proposed using such effects to infer the mass of ULDM.
For intermediate-mass black holes (IMBHs) with masses of $ \sim 10^2 - 10^5 M_{\odot}$, a different handle becomes available. If a solar-mass compact object is captured by an IMBH, the gravitational waves (GWs) emitted from the inspiral can fall within the sensitivity of the Laser Interferometer Space Antenna (LISA)~\cite{LISA:2022kgy,LISA:2017pwj, eLISA:2013xep, Barack:2003fp}. The presence of the boson star environment modifies the GW waveform, thereby enabling measurements of the underlying ULDM parameters~\cite{Macedo:2013qea,Huang:2018pbu,Hook:2017psm,Cole:2022yzw,Kadota:2023wlm,Boudon:2023vzl,Aurrekoetxea:2023jwk, Aurrekoetxea:2024cqd,Croon:2017zcu,Choi:2018axi,Cao:2024wby,Blas:2024duy,Takahashi:2024fyq,Kim:2024rgf,Roy:2025qaa,Berezhiani:2023vlo}.
Analogous environmental imprints have been extensively studied in the context of collisionless DM~\cite{Eda:2013gg,Barausse:2014tra,Eda:2014kra,Yue:2017iwc,Bertone:2019irm,Cardoso:2019rou,Hannuksela:2019vip,Kavanagh:2020cfn,Coogan:2021uqv,Cole:2022fir,Cardoso:2019rou,Yang:2025yej}, and self-interacting DM~\cite{Banik:2025fnc}.

In both mass regimes, the central black hole reshapes the density of the ULDM condensate, thereby strongly impacting phenomenological conclusions. This renders a comprehensive study of BS-BH systems necessary for robust phenomenology. In this work, we investigate such a system, solving the governing Gross-Pitaevskii-Poisson (GPP) equation~\cite{gross1963hydrodynamics,pitaevskii1959properties,pitaevskii2016bose,Chavanis:2011zi,Eby:2015hsq, Schiappacasse:2017ham} to determine the density profile of the boson star hosting the black hole. We show that the presence of a central black hole modifies the boundary conditions, which we consistently take into account to obtain the equilibrium configurations of the BS-BH system. We first show that the presence of a black hole enhances the central density of the boson star, while reducing its size. In our numerical approach, we explore the full parameter space of such a system in the nonrelativistic limit, carefully differentiating between stable and unstable configurations. We find that unstable configurations arise only for attractive interactions of ULDM, placing a bound on the minimum possible value of the self-coupling and the maximum mass of the boson star, which depend on the mass ratio between the central black hole and the boson star. We present our results in terms of dimensionless quantities, thereby being agnostic to the precise values of the boson star parameters and the black hole mass. This allows one to obtain the physical quantities of interest, specifically, the central density of the boson star and its size.

In addition, we consider an approximate form for the density profile of the system, an \textit{ansatz}, for analytical insight, compatible with the boundary conditions for the numerical solution of the GPP equation. Using this ansatz, we obtain, for example, the modified mass-radius relation for the boson star, which we compare to our numerical results, finding close agreement, in particular for attractive interactions and for repulsive interactions when the mass ratio between the black hole and boson star is $\lesssim 0.3$.

As a phenomenological application, we investigate how the newly computed BS-BH system density profiles modify the GW phase during IMBH inspirals. Using a dynamical friction (DF) force that accounts for the quantum pressure of ULDM~\cite{Lancaster:2019mde}, we demonstrate that this effect significantly reduces the energy lost due to DF, thereby greatly impacting the orbital evolution.
Accordingly, we map out the ULDM parameter space that LISA can probe based on a Fisher information matrix analysis.

Our paper is structured as follows. Section~\ref{sec:density_profile} introduces our model of ULDM with quartic self-interactions and derives the
equations of hydrostatic equilibrium to describe the gravitating condensate in the nonrelativistic limit, i.e., the GPP equations. We solve these equations numerically to obtain the density profiles of the BS-BH system and explore the parameter space of the model. In Sec.~\ref{sec:ansatz}, we consider an approximate form for the density profile, motivated by our numerical analyses. We study various properties based on this ansatz for the density profile, comparing against the numerical results. Based on the available parameter space of BS-BH systems, we investigate in Sec.~\ref{sec:gw_probe} the region in the plane of coupling and ULDM mass that can be probed by LISA, based on GW dephasing when a secondary black hole merges with this system. Finally, we conclude in Sec.~\ref{sec:conclude}.

\section{Boson star density profile around black holes}
\label{sec:density_profile}

We consider a real scalar field $\phi$ charged under a $\mathbb{Z}_2$ symmetry, with the following action
\begin{equation}
    S_{\phi} = \int d^4x \sqrt{|g|}\,\left\{\frac{1}{2}g^{\mu\nu}\partial_{\mu}\phi\partial_{\nu}\phi-\frac{m^2}{2}\phi^2-\frac{\lambda}{4!}\phi^4\right\}\,,
    \label{eq:scalar_action}
\end{equation}
where we include terms that are up to renormalizable order.%
\footnote{Unless otherwise stated, we adopt natural units $\hbar = c = 1$ for notational brevity. Throughout, we use the mostly minus metric convention such that the Minkowski metric reads $\eta_{\mu \nu} \equiv {\rm diag}(1,-1,-1,-1)$.}
Here, $m$ is the mass of the scalar field and $\lambda$ gives the strength of the quartic self-interaction, with $\lambda < 0 \,(\lambda >0)$ denoting attractive (repulsive) interactions.
We consider the metric $g_{\mu\nu}$ from,
\begin{equation}
    ds^2 = (1+2\Phi)dt^2 -(1-2\Phi)\delta_{ij}dx^idx^j\,.
    \label{eq:line_el}
\end{equation}
Here $\Phi$ denotes the metric perturbation, which is sourced by the self-gravitating condensate, which we call a boson star, and the black hole hosted at its center. Working in the nonrelativistic limit, where the momentum of the particles $|\vec{p}| \ll m$, we can express the real scalar field in terms of a complex scalar field $\psi$ and a phase,
\begin{equation}
    \phi = \frac{1}{\sqrt{2m}}\left(e^{-imt}\psi+e^{imt}\psi^*\right)\,.
    \label{eq:nonrel_osc}
\end{equation}
In what follows, we neglect gradients $|\nabla \psi/\psi| \ll m$ and assume a slowly varying metric perturbation $|\dot \Phi/\Phi| \ll m$. Given that oscillatory terms $\propto e^{-imt}$ that average out to zero over sufficiently long timescales, we also neglect these. Then, substituting \eqref{eq:nonrel_osc} and the metric derived from \eqref{eq:line_el} into \eqref{eq:scalar_action}, we obtain
\begin{equation}
    \begin{aligned}
        S_{\phi}
        &= \int d^4x \bigg\{\frac{i}{2}\left(\dot{\psi}\psi^{*}-\psi\dot{\psi}^{*}\right)-\frac{\partial_i\psi\,\partial^i\psi^{*}}{2m} \\
        &\quad - \frac{m}{2}\Phi|\psi|^2 -\frac{\lambda\,(|\psi|^2)^2}{16m^2}\bigg\}\,\,,
    \end{aligned}
    \label{eq:action_non-rel}
\end{equation}
where $\cdot \equiv d/dt$. This allows us to derive the Gross-Pitaevskii equation~\cite{Chavanis:2011zi, Chavanis:2011zm, Eby:2015hsq, Schiappacasse:2017ham} for $\psi$
\begin{equation}
    i\dot{\psi} = -\frac{1}{2m}\nabla^2\psi + m\psi\left(\Phi+\frac{\lambda}{8m^3}|\psi|^2\right)\,,
    \label{eq:GNP}
\end{equation}
which is the nonrelativistic limit of the Klein-Gordon equation.

As we are interested in scalar masses around $m \sim 10^{-20}\text{--}10^{-14}~\text{eV}$, the high occupation number allows the bosons to condense into their lowest momentum state and behave as a single macroscopic fluid. We then proceed with a mean-field approximation and decompose the condensate using a Madelung transformation \cite{madelung1927quantum}
\begin{equation}
    \psi(\vec{r},t) \equiv |\psi(\vec{r},t)|e^{iS(\vec{r},t)}=\sqrt{\frac{\rho(\vec{r},t)}{m}}\,e^{iS(\vec{r},t)}\,,
    \label{eq:madelung_transform}
\end{equation}
which we substitute into \eqref{eq:GNP}. Separating out the real and imaginary parts yields
\begin{subequations}
    \begin{align}
        \dot{\rho}+\vec{\nabla}\cdot(\rho\,\vec{u}) &= 0\,,
        \label{eq:euler1}\\
        \dot{\vec{u}}+(\vec{u}\cdot\vec{\nabla})\vec{u} &= -\vec{\nabla}\left[\Phi+\frac{\lambda}{8m^4}\rho-\frac{\nabla^2\sqrt{\rho}}{2m^2\sqrt{\rho}}\right]\,,
        \label{eq:euler2}
    \end{align}
\end{subequations}
where we have defined the fluid velocity $\vec{u} \equiv \vec{\nabla} S/m$. In what follows, we will focus on the case of hydrostatic equilibrium, where $\vec{u} = 0$, and the solutions are time independent. The total boson star mass, $M$, is defined from its density profile $\rho(\vec{r})$,
\begin{equation}
    M = \int \rho(\vec{r})\,d^3r\,.
    \label{eq:mass_sol}
\end{equation}
Einstein's equations in this nonrelativistic limit\footnote{For studies considering relativistic treatment, we refer readers to Refs.~\cite{Barranco:2012qs,Sadeghian:2013laa,Herdeiro:2014goa,Barranco:2017aes,
Feng:2021qkj,Glavan:2025khe,DeLuca:2023laa,Alcubierre:2024mtq}.} reduce to the Poisson equation for the metric perturbation $\Phi$, which is sourced from the condensate~\cite{Boehmer:2007um, Chavanis:2011zi, Chavanis:2011zm} and the black hole~\cite{Chavanis:2019bnu}
\begin{equation}
    \nabla^2\Phi = 4\pi G\,(\rho+\rho_{\rm BH})\,,
    \label{eq:poisson}
\end{equation}
where $G$ is Newton's constant and the black hole density reads as $\rho_{\rm BH} \equiv M_{\rm BH}\,\delta^{(3)}(\vec{r})$. The solution to \eqref{eq:poisson} is readily obtained as
\begin{equation}
    \Phi(r) = -G\int \frac{\rho(\vec{r}^{\prime})}{|\vec{r} - \vec{r}^{\prime}|} d^3r^{\prime}- \frac{GM_{\rm BH}}{r}\,.
    \label{eq:poisson_sol}
\end{equation}
For hydrostatic equilibrium, \eqref{eq:euler2} reduces to
\begin{equation}
    \Phi +\frac{\lambda}{8m^4}\rho - \frac{\nabla^2\sqrt{\rho}}{2m^2\sqrt{\rho}} = \epsilon \equiv {\rm const.}\,,
    \label{eq:eigenenergy}
\end{equation}
where $\epsilon$ represents the eigenenergy density of the system.

Specifically, we will study the spherically symmetric solutions for the density profile. To this end, we take a Laplacian of \eqref{eq:eigenenergy} and substitute \eqref{eq:poisson} to obtain,
\begin{equation}
    \nabla^2\left(\frac{\nabla^2\sqrt{\rho}}{2m^2\sqrt{\rho}}\right) - \frac{\lambda}{8m^4}\nabla^2\rho = 4\pi G(\rho+\rho_{\rm BH})\,.
    \label{eq:hydro}
\end{equation}
Equation~\eqref{eq:hydro} represents the equilibrium system formed from the balance between the quantum pressure, the pressure induced due to self-interactions of the boson star and the gravitational attraction of the condensate and the black hole. A similar equation was derived in Ref.~\cite{Chavanis:2019bnu}, focusing on an analytic treatment of the same based on a Gaussian approximation for the density profile. This approximation depended only on a single parameter, the characteristic radius of the boson star. However, as we will show, in our numeric approach, the external potential sourced by the black hole directly modifies the nature of the exact solution. Consequently, the analytic approximation should be modified, which we discuss in Sec.~\ref{sec:ansatz}.

To progress, we closely follow Ref.~\cite{Chavanis:2011zm} and introduce the following length scale:
\begin{equation}
    b \equiv \frac{1}{2G M m^2}\,,
    \label{eq:length_scale_b}
\end{equation}
which can be interpreted as the radius of a gravitating boson star without self-interactions \cite{Ruffini:1969qy}.
We then rescale to a dimensionless variable $\vec{x} \equiv \vec{r}/b$ and accordingly define the dimensionless densities,
\begin{subequations}
    \begin{align}
        n(x) &\equiv \frac{4\pi b^3}{M}\,\rho(x)\,,
        \label{eq:scaled_density} \\
        n_{\rm BH}(x) &\equiv 4\pi b^3 \frac{M_{\rm BH}}{M}\delta^{(3)}(\vec{x})\,,
    \end{align}
\end{subequations}
along with a dimensionless parameter $\chi$,
\begin{equation}
    \chi \equiv \frac{GM^2\lambda}{8\pi} \,,
    \label{eq:chi}
\end{equation}
which captures the dependence on the boson star parameters $(M,\lambda)$. The normalization of the mass \eqref{eq:mass_sol} reads now
\begin{equation}
    \int_0^{\infty} dx \,x^2n(x)= 1\,.
\end{equation}
Finally, Eq.~\eqref{eq:hydro} becomes
\begin{equation}
    \tilde{\nabla}^2\left(\frac{\tilde{\nabla}^2\sqrt{n}}{\sqrt{n}}\right) - \chi\tilde{\nabla}^2n = (n+n_{\rm BH})\,.
    \label{eq:red_hydro}
\end{equation}
Equation~\eqref{eq:red_hydro} is a fourth-order, nonlinear differential equation, requiring four boundary conditions to solve. In what follows, we will consider density profiles that are nonsingular at the center $x = 0$. This allows us to consider a Taylor expansion of the form~\cite{Membrado:1989bqo}
\begin{equation}
    n(x) \approx n_0 + n_1 x + \frac{n_2\,x^2}{2} + \frac{n_3\, x^3}{6}+\mathcal{O}(x^4)\,.
    \label{eq:expansion}
\end{equation}
Since $n(0) \equiv n_0$, this coefficient is associated with the central density of the boson star.
We note that the incorporation of the quantum pressure term, corresponding to the first term of the lhs of \eqref{eq:red_hydro}, allows us to obtain this class of solutions, which is nonsingular at $x = 0$. Stable solutions, neglecting the quantum pressure, can be obtained for repulsive interactions, but result in the profile diverging for $r \to 0$, as shown in Refs.~\cite{Chavanis:2019bnu, DeLuca:2023laa}.

Next, we plug \eqref{eq:expansion} into \eqref{eq:red_hydro}, to obtain
\begin{equation}
    n_3 = \frac{n_1}{2n_0^2} \left(5n_0n_2-3n_1^2\right) + \chi n_0 n_1\,.
    \label{eq:n3}
\end{equation}
Thus, the constant $n_3$, related to the third derivative of the density profile, is determined by a combination of the other lower derivatives of the density profile at $x = 0$.

Performing a volume integral on both sides of \eqref{eq:red_hydro} and applying Gauss's divergence theorem on the lhs yields
\begin{equation}
    x^2\left[-\frac{d}{dx}\left(\frac{\tilde{\nabla}^2\sqrt{n}}{\sqrt{n}}\right)-\chi\frac{dn}{dx}\right] = \frac{M(x)}{M}+\frac{M_{\rm BH}}{M}\,.
\end{equation}
We again substitute the expansion \eqref{eq:expansion} and take the limit $x \to 0$ to obtain
\begin{equation}
    \frac{n_1}{n_0}= -\frac{M_{\rm BH}}{M} \equiv - \kappa\,.
    \label{eq:n1}
\end{equation}
The first derivative of the profile at $x = 0$ hence is determined by the ratio between the masses of the black hole and the boson star, which we denote by $\kappa$. Therefore, the gravitational potential of the black hole intrinsically alters the boundary conditions, an aspect that has not been explicitly discussed by earlier works, such as Refs.~\cite{Chakrabarti:2022owq,Dave:2023wjq}.

The physical interpretation of $n_2$ is obtained by studying \eqref{eq:eigenenergy}, which, after rescaling, becomes,
\begin{equation}
    \tilde{\Phi} + \chi n - \left[\frac{d^2n}{dx^2}+\frac{1}{xn}\frac{dn}{dx}-\frac{1}{n^2}\left(\frac{dn}{dx}\right)\right] = \tilde{\epsilon} \,,
    \label{eq:poisson_red}
\end{equation}
where $\tilde{\Phi} \equiv \Phi \,b/(GM)$ and $\tilde{\epsilon} \equiv \epsilon \,b/(GM)$. On inserting \eqref{eq:expansion} and \eqref{eq:poisson_sol} into \eqref{eq:poisson_red}, and taking the limit $x \to 0$, we obtain
\begin{equation}
    -\int_0^{\infty}x\,n(x)\,dx + \chi n_0 - \frac{n_2}{n_0} + \frac{n_1^2}{4n_0^2} = \tilde{\epsilon}\,.
\end{equation}
This implies that the constant $n_2$ relates to the energy of the system.

\begin{figure}[htbp]
    \centering
    \includegraphics[width=0.48\textwidth]{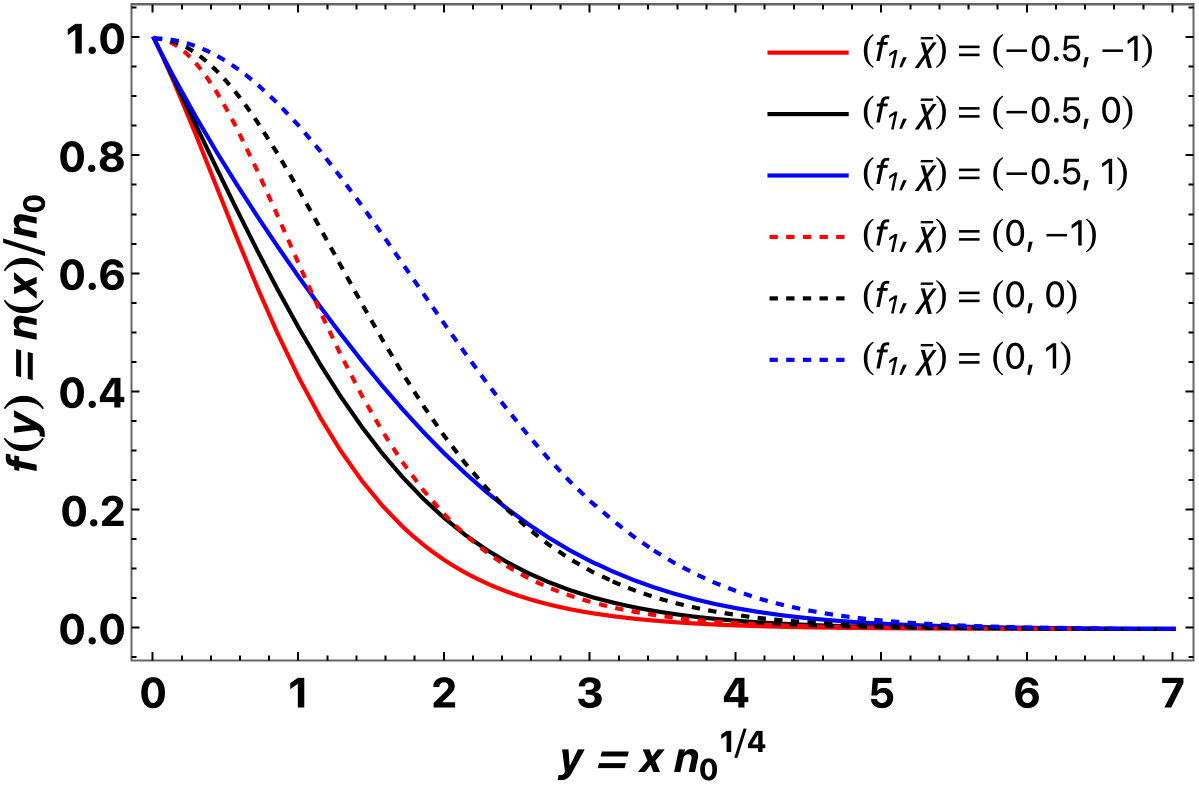}
    \caption{Examples of the solutions to the differential equation \eqref{eq:red_hydro2}, where $y$ is the scaled radius. The quantity $\bar{\chi}$ corresponds to the scaled interaction strength defined from \eqref{eq:chi-bar}. The nonzero value of $f_1$ also evidently modifies the slope of the profiles. All of these colored lines meet at $y = 0$ as they share the same boundary conditions: $f(0) = 1$ and $f'(0) = f_1$.
    }
    \label{fig:solutionEg}
\end{figure}

For the numerical approach we adopt, we apply a final rescaling to \eqref{eq:red_hydro} defined as,
\begin{equation}
    \begin{aligned}
 &f(x) \equiv \frac{n(x)}{n_0}\,,\quad f_{\rm BH}(x)\equiv \frac{n_{\rm BH}(x)}{n_0}\\
 &y \equiv n_0^{1/4}x\,,\quad \bar{\chi} \equiv n_0^{1/2}\chi\,,
 \label{eq:chi-bar}
    \end{aligned}
\end{equation}
thereby normalizing the central density to unity. We then have the following differential equation to solve,
\begin{equation}
    \begin{aligned}
 &f''''+\frac{4}{y}f''' - 10\frac{f'f''}{yf}+6\frac{f'^3}{yf^2}-3\frac{f'''f'}{f}-2\frac{f''^2}{f} \\
 &\qquad +7\frac{f'^2f''}{f^2}-3\frac{f'^4}{f^3}-2\bar{\chi} f\left(f''+2\frac{f'}{y}\right) \\
 & \qquad =2f(f+f_{\rm BH})\,,
    \end{aligned}
    \label{eq:red_hydro2}
\end{equation}
where $' \equiv d/dy$. The boundary conditions we consider to solve this equation are determined by appropriately rescaling the coefficients $n_i$,
\begin{subequations}
    \begin{align}
        f(0) &= \frac{n_0}{n_0} = 1\,, \\
        f'(0) &= \frac{n_1}{n_0^{5/4}} = -\kappa \,n_0^{-1/4} \equiv f_1\,,
        \label{eq:f1}
        \\
        f''(0) &= \frac{n_2}{n_0^{6/4}}\equiv f_2\,,
        \\
        f'''(0) &= \frac{n_3}{n_0^{7/4}} = \frac{f_1}{2}\left(5f_2-3f_1^2\right) + \bar{\chi} f_1\equiv f_3\,.
    \end{align}
    \label{eq:boundary_conditions}
\end{subequations}
As we solve for $y >0$, $f_{\rm BH} = 0$ and the black hole contribution enters through the boundary condition $f_1$. To solve \eqref{eq:red_hydro2}, we specify $f_1$ and $\bar{\chi}$, with the unknown $f_2$ determined using the shooting method, with the requirement that $f(y)$ monotonically decreases to zero as $y \to \infty$. Therefore, the solution, specified by this $f_2$ depends on the parameter set $(f_1, \bar{\chi})$.
Accordingly, we solve \eqref{eq:red_hydro2} for several sets of $(f_1,\bar{\chi})$ from $\bar{\chi} \in \{-3,3\}$ and $f_1 \in \{-1.7,0\}$.
In Fig.~\ref{fig:solutionEg}, we show examples of such solutions. For fixed $f_1$, as $\bar{\chi}$ increases from $-1$ to $1$, the scaled profile spreads out. This is due to effectively changing the interaction type from attractive to noninteracting and finally repulsive. For fixed $\bar{\chi}$, a nonzero $f_1$ always pulls the profile inward, due to the additional gravitational potential introduced by the black hole.

\begin{figure}[htbp]
    \centering
    \includegraphics[width=0.482\textwidth]{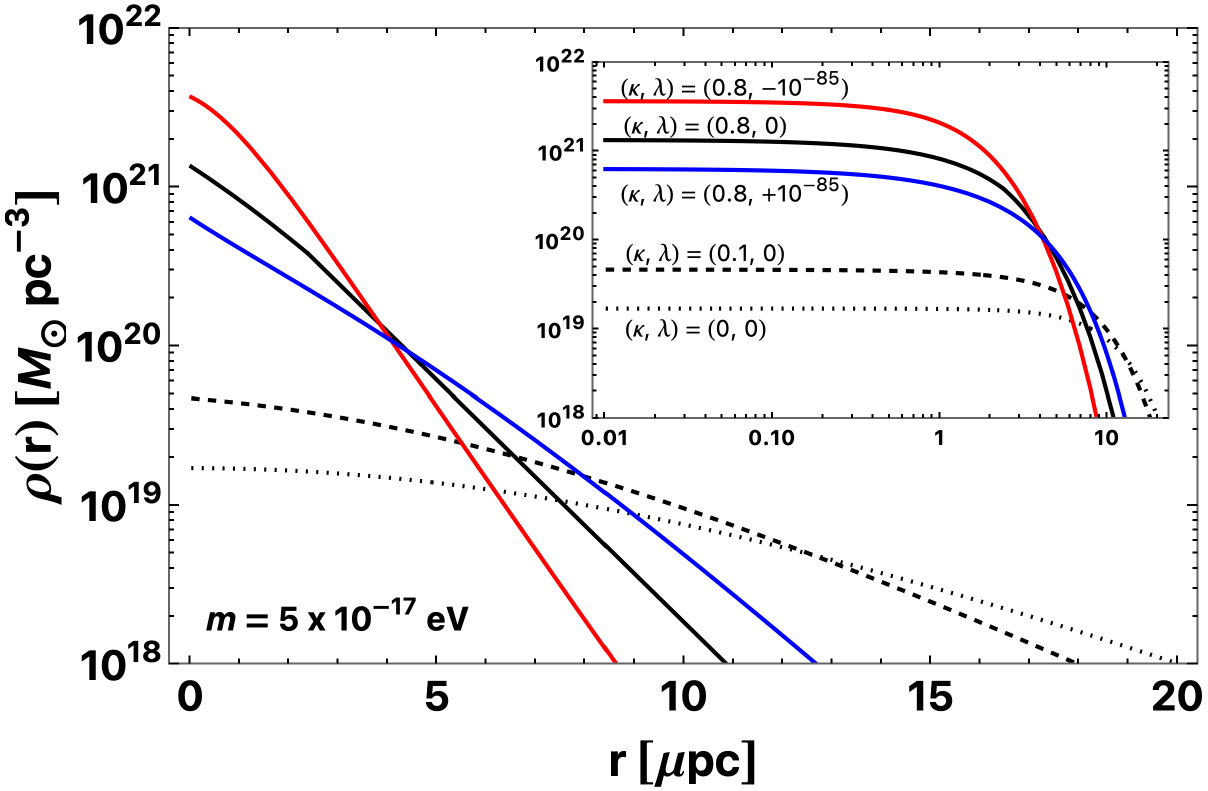}
    \caption{Examples of density profiles with fixed DM mass $m =5\times10^{-17}$~eV.
        The dotted line indicates a gravitating boson star $(\lambda = 0)$ with no central black hole, of size $\approx 2 \times 10^{-5}$~pc, resulting in a boson star of mass $\sim 1.5\times 10^5\,M_{\odot}$. The dashed and solid lines indicate that the boson star hosts a central black hole of mass $M_{\rm BH} = 10^5\,M_{\odot}$, with the boson star mass given by $M = M_{\rm BH}/\kappa$. Red and blue lines indicate attractive and repulsive interactions, respectively. The central black hole reduces the size of the boson star and increases the density of the boson star. Repulsive interactions $(\lambda >0)$ reduce the density, whereas attractive interactions $(\lambda < 0)$ enhance it. \textit{Inset}: Same density profiles but in log-log scale.
    }
    \label{fig:density_profile_eg}
\end{figure}

To obtain the physical density profile, one revisits the mass normalization, which now reads,
\begin{equation}
    \int_0^{\infty}dy\,y^2f(y)=n_0^{-1/4}\,.
    \label{eq:n0}
\end{equation}
Having determined the profile $f(y)$ for a given $(f_1,\bar{\chi})$, this equation allows us to determine the normalization constant $n_0$. We then obtain the true ratio between the black hole and boson star masses, $\kappa$, from \eqref{eq:f1}. This gives the density profile of the boson star hosting a black hole through \eqref{eq:scaled_density},
depending on the two parameters $\chi$ and $\kappa$ defined in \eqref{eq:chi} and \eqref{eq:n1}, respectively. The latter two are derived from the combination of model parameters $(m,\lambda,M,M_{\rm BH})$, which, when specified, uniquely determine the density profile. Examples of the density profiles are given in Fig.~\ref{fig:density_profile_eg} for fixed black hole and DM masses. Increasing the ratio $\kappa$ (or reducing the boson star mass $M$) increases the central density, while shrinking the radius of the boson star. On the other hand, attractive (repulsive) interactions result in a larger (smaller) central density and a more (less) compact boson star.
Finally, we note that incorporating relativistic effects can result in a larger central density, as discussed in Appendix~\ref{app:comparison}.

In this manner, we solve \eqref{eq:red_hydro2} to obtain the density profiles for the BS-BH system, i.e., the equilibrium solutions. However, not any value of parameters $(m,\lambda,M,M_{\rm BH})$ can lead to a stable configuration.
Increasing the absolute value of a negative $\lambda$ implies an increase in strength of attractive self-interactions. Therefore, in this case, if $|\lambda|$ is too large, the outward quantum pressure cannot balance the inward gravity and attractive self-interaction, causing the system to collapse. Therefore there is a lower bound $\lambda_*$ for a stable configuration. For a critical system with $\lambda_*$, increasing $M_\BH$ will increase the gravitational potential and make the system collapse, therefore $|\lambda_*|$ is smaller for larger $M_\BH$.
For attractive self-interactions, there is an upper bound for boson star mass $M_*$, above which the self-gravity is too large. Since larger $|\lambda|$ and larger $M_\BH$ give larger inward force, $M_*$ is correspondingly smaller for larger $|\lambda|$ and $M_\BH$. We quantify these inferences in the forthcoming subsections. Essentially, the stable equilibrium solutions correspond to minima of the energy of the system, a point we revisit Sec.~\ref{sec:ansatz}. Note that for a noninteracting boson star or one with repulsive interactions, we find all configurations to be stable.\footnote{When relativistic effects are taken into account, a boson star without interactions or with repulsive interactions exhibits a maximum mass, see for example Ref.~\cite{Liebling:2012fv}. We expect such a limit to exist for the BS-BH system when a full relativistic treatment is considered, but this is not the purpose of this present study, where we restrict ourselves to the nonrelativistic regime.}

We will now investigate various properties of interest of the BS-BH system in the space of the model parameters. To this end, we keep the DM mass $m$ fixed as this serves to set the scale of the system. This model parameter also does not appear explicitly in the definition of the $\chi$ defined in \eqref{eq:chi}, one of the control parameters for the differential equation \eqref{eq:red_hydro}. Then, we study properties of phenomenological interest, such as the central density and the actual size of the system in the space of the three remaining parameters. To streamline our discussion, we consider these properties in the plane of two parameters, keeping the third fixed. It is convenient to define the scattering length of the interaction,
\begin{equation}
    a \equiv \frac{\lambda}{32\pi m}\,.
    \label{eq:scatt_length}
\end{equation}
We then study the aforementioned properties of the system in the plane of $(M, M_{\rm BH})$ with the scattering length $a$ fixed, and in the plane of $(a, M_{\rm BH})$ with the total mass of the boson star $M$ being fixed.

\subsection{Fixed magnitude of the coupling}
\label{subsec:fix_a}

\begin{figure*}[htbp]
    \centering
    \includegraphics[width=0.47\textwidth]{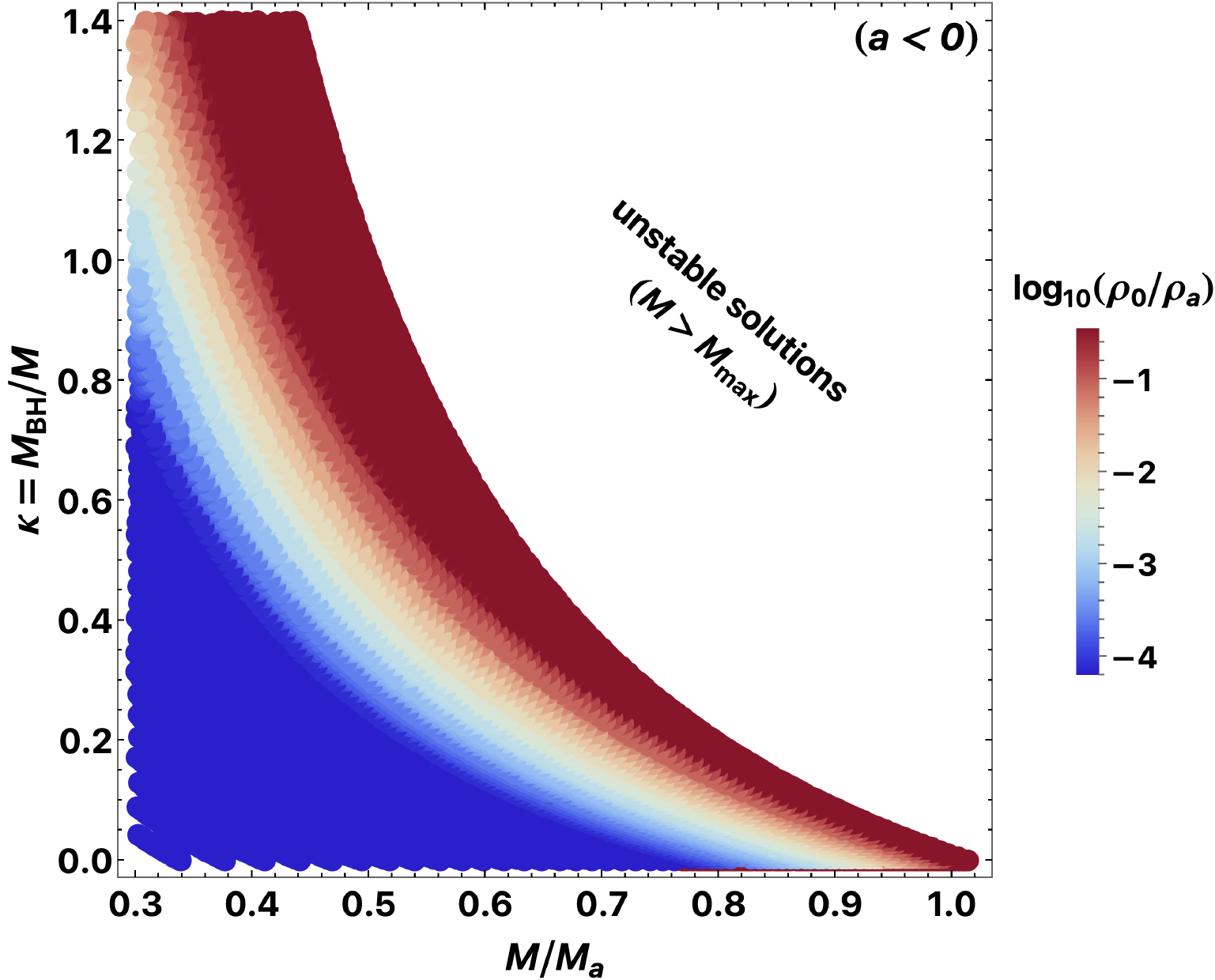}\,\,
    \includegraphics[width=0.47\textwidth]{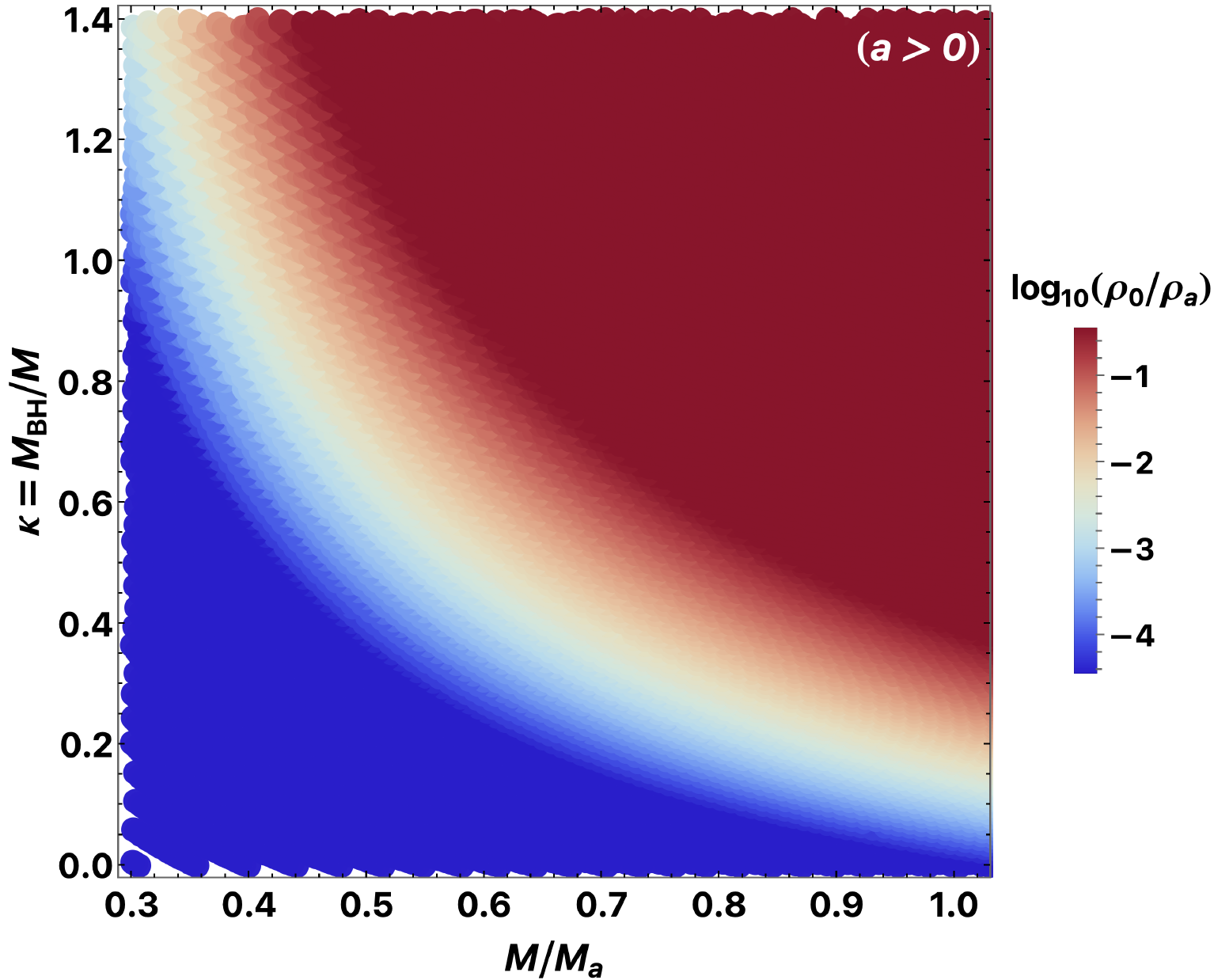}\\[2mm]
    \includegraphics[width=0.47\textwidth]{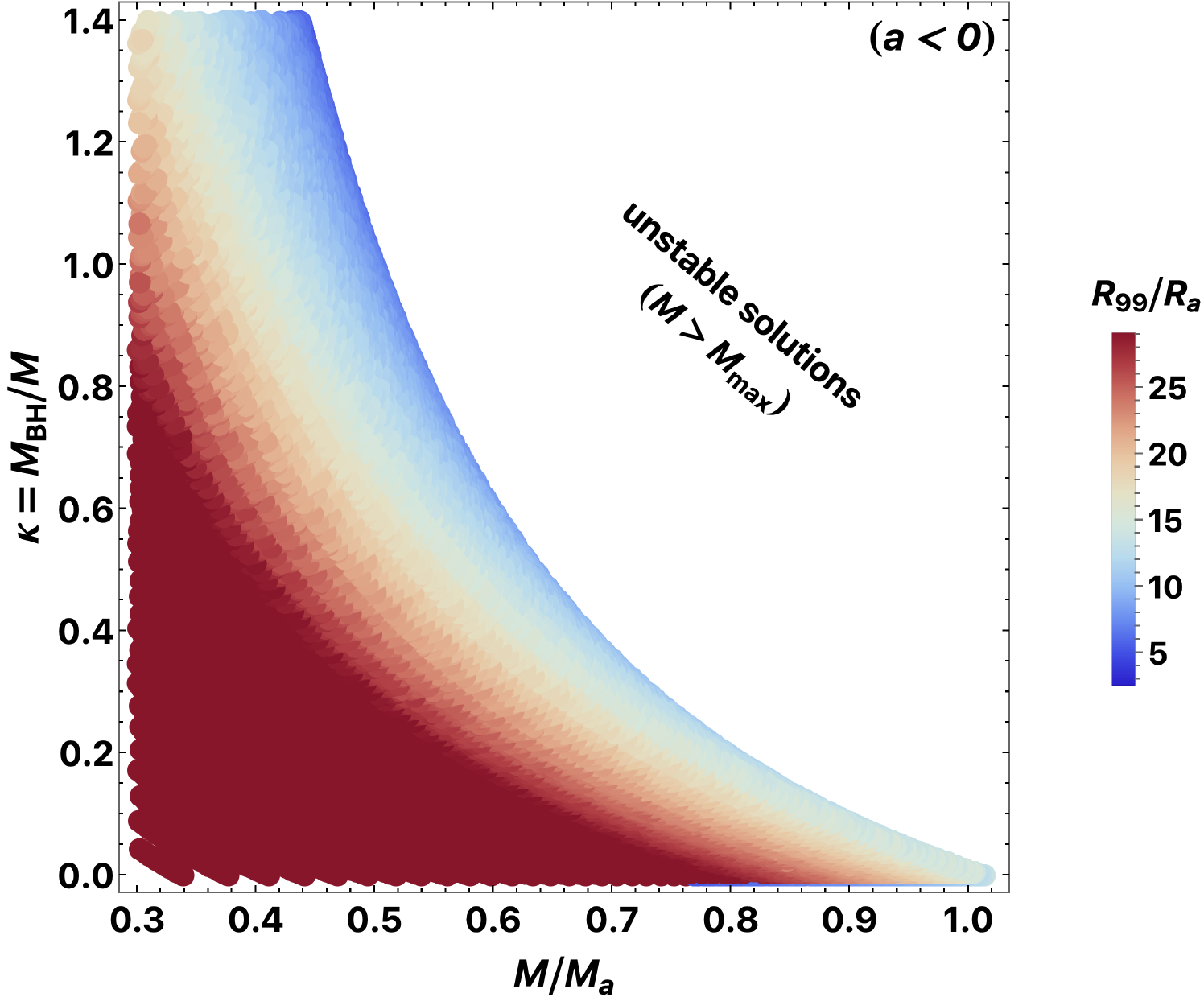}\,\,
    \includegraphics[width=0.47\textwidth]{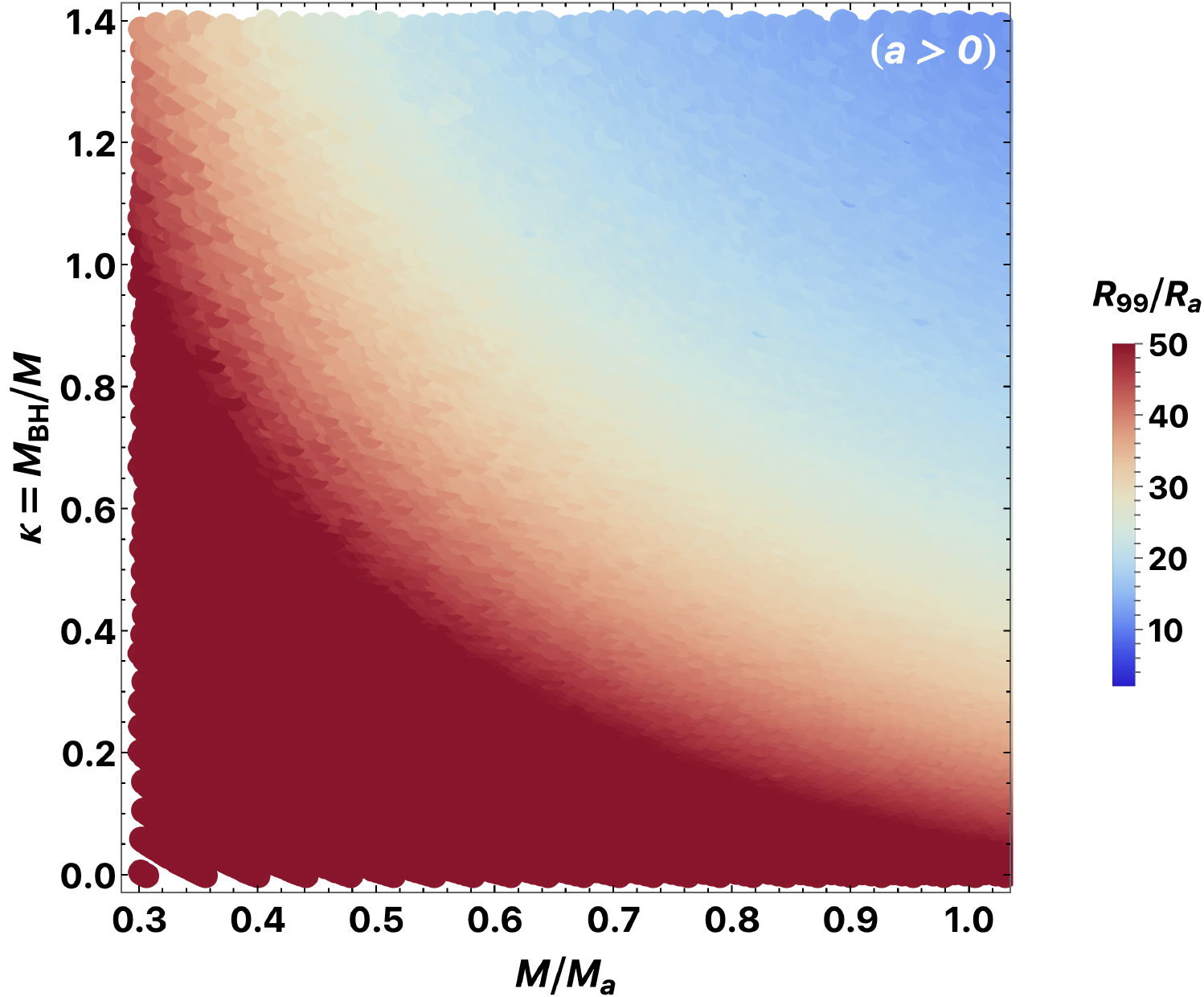}
    \caption{Central density (\textit{top}) normalized to $\rho_a\equiv Gm^4/|a|^2$ and radius containing $99\%$ of the mass (\textit{bottom}) of boson stars hosting central black holes, normalized to $R_a \equiv \sqrt{|a|/(Gm^3)}$. We show scenarios with attractive (\textit{left column}) and repulsive (\textit{right column}) interactions in the plane of the mass of the boson star, normalized by $M_a \equiv (Gm|a|)^{-1/2}$ and the ratio $\kappa$, for fixed magnitude of the interaction strength. For attractive-type interactions $(a < 0)$, there exists a maximum allowed mass $M_{\rm max}$, for fixed $\kappa$, beyond which the system becomes unstable. No such restriction arises for repulsive-type interactions, implying the full parameter space is allowed.
    }
    \label{fig:afix_att_rep}
\end{figure*}

In this case, to fix the DM mass and the scattering length (corresponding to fixing the quartic coupling $\lambda$), we define the following length and mass scales for normalization:
\begin{align}
    R_a \equiv \left(\frac{|a|}{Gm^3}\right)^{1/2}\,,\quad \quad
    M_a \equiv \frac{1}{\sqrt{Gm|a|}}\,.
    \label{eq:norm_fixa}
\end{align}
The quantity $R_a$ can be discerned from \eqref{eq:hydro}, being the length scale at which the self-interaction pressure [which scales as $\sim (|a|/m^3)(1/r^2)(M/r^3)$ on dimensional grounds] becomes comparable to the gravitational force of the boson star (scaling as $\sim G M/r^3$), when the quantum pressure is neglected. This corresponds to the parameter $\chi \sim 1$ in \eqref{eq:red_hydro}, leading to the mass scale $M_a$. One can therefore define a density from these quantities for normalization:
\begin{equation}
    \rho_a \equiv \frac{M_a}{R^3_a} = \frac{Gm^4}{|a|^2}\,.
\end{equation}
This allows us to define the following dimensionless boson star mass, central density and radius containing $99\%$ of the mass of the boson star, respectively, as,
\begin{subequations}
    \begin{align}
        \frac{M}{M_a} &= \frac{\sqrt{|\chi|}}{2} = \frac{\sqrt{|\bar{\chi}|}}{2n_0^{1/4}}\,, \\
        \frac{\rho_0}{\rho_a} &= \frac{|\chi|^2}{8\pi}n_0 = \frac{|\bar{\chi}|^2}{8\pi}\,, \\
        \frac{R_{99}}{R_a} &= \frac{x_{99}}{\sqrt{|\chi|}} = \frac{y_{99}}{\sqrt{|\bar{\chi}|}}\,,
    \end{align}
\end{subequations}
where we have explicitly shown the conversions to extract the physical quantities from the various rescalings applied to the differential equation \eqref{eq:hydro}. We have defined the central density $\rho_0 \equiv \rho(0)$.

In Fig.~\ref{fig:afix_att_rep}, we show the dependence of $\rho_0$ and $R_{99}$ in the plane of the remaining free parameters $(M,M_{\rm BH})$. Although we have fixed the value of the scattering length, essentially fixing the coupling $\lambda$, its sign in \eqref{eq:scalar_action} leads to different behavior, as previously mentioned. In particular, for $a < 0$, we observe that there exists a maximum mass of the boson star $M_{\rm max}$ for a fixed value of the ratio $\kappa$, for which the system remains stable. Increasing $\kappa$ causes the $M_{\rm max}$ to decrease, due to the increased gravitational potential of the black hole.
We note that this existence of $M_{\rm max}$ for attractive self-interactions in the absence of a central black hole was identified in \cite{Chavanis:2011zi,Chavanis:2011zm}, and the decreasing behavior of $M_{\rm max}$ with the mass of the black hole was also studied in \cite{Chavanis:2019bnu}.
However, for repulsive interactions, $a > 0$, there exists no such restriction on the mass of the boson star, i.e., all configurations are allowed. Inherently common to both types of interactions is that the central density of the boson star increases and its size (radius) decreases, if one increases either or both of $(M,\kappa)$, which can be seen from the transition from blue to red shades in Fig.~\ref{fig:afix_att_rep}. Boson stars with repulsive interactions tend to have a larger extent and lower density than their counterparts with attractive interactions.

\subsection{Fixed boson star mass}
\label{subsec:fix_M}

In this alternative parametrization of the system, to fix the mass of the constituent DM $m$ and the total boson star mass $M$, we define the following scales for the scattering length and the size of the boson star,
\begin{align}
    R_M \equiv \frac{1}{GM\,m^2} = 2b\,,\quad \quad
    a_M \equiv \frac{1}{GM^2\,m}\,.
    \label{eq:norm_fixM}
\end{align}
We note that this parametrization is well suited for more particle physics-oriented phenomenological applications, and we make use of this in Sec.~\ref{sec:gw_probe}.

Physically, $R_M$ corresponds to the typical size of a gravitationally stable boson star with no interactions, and is related to the length-scale $b$, which we have used to obtain \eqref{eq:red_hydro}. The interaction length-scale $a_M$ again corresponds to the regime where the self-interactions become comparable to the attractive gravitational force, meaning $\chi \sim 1$ for fixed boson star and DM mass. We can then derive the scaling density for normalization,
\begin{equation}
    \rho_M \equiv \frac{M}{R^3_M} = G^3 M^4m^6\,.
    \label{eq:rhoM}
\end{equation}
Then, the dimensionless scattering length, central density and radius containing $99\%$ of the mass of the boson star are, respectively, given by,
\begin{subequations}
    \begin{align}
        \frac{a}{a_M} &= \frac{\chi}{4} = \frac{\bar{\chi}}{4n_0^{1/2}}\,, \\
        \frac{\rho_0}{\rho_M} &= \frac{2}{\pi}n_0 \,, \\
        \frac{R_{99}}{R_M} &= \frac{x_{99}}{2} = \frac{y_{99}}{2n_0^{1/4}} \,.
    \end{align}
\end{subequations}

\begin{figure}[htbp]
    \centering
    \includegraphics[width=0.47\textwidth]{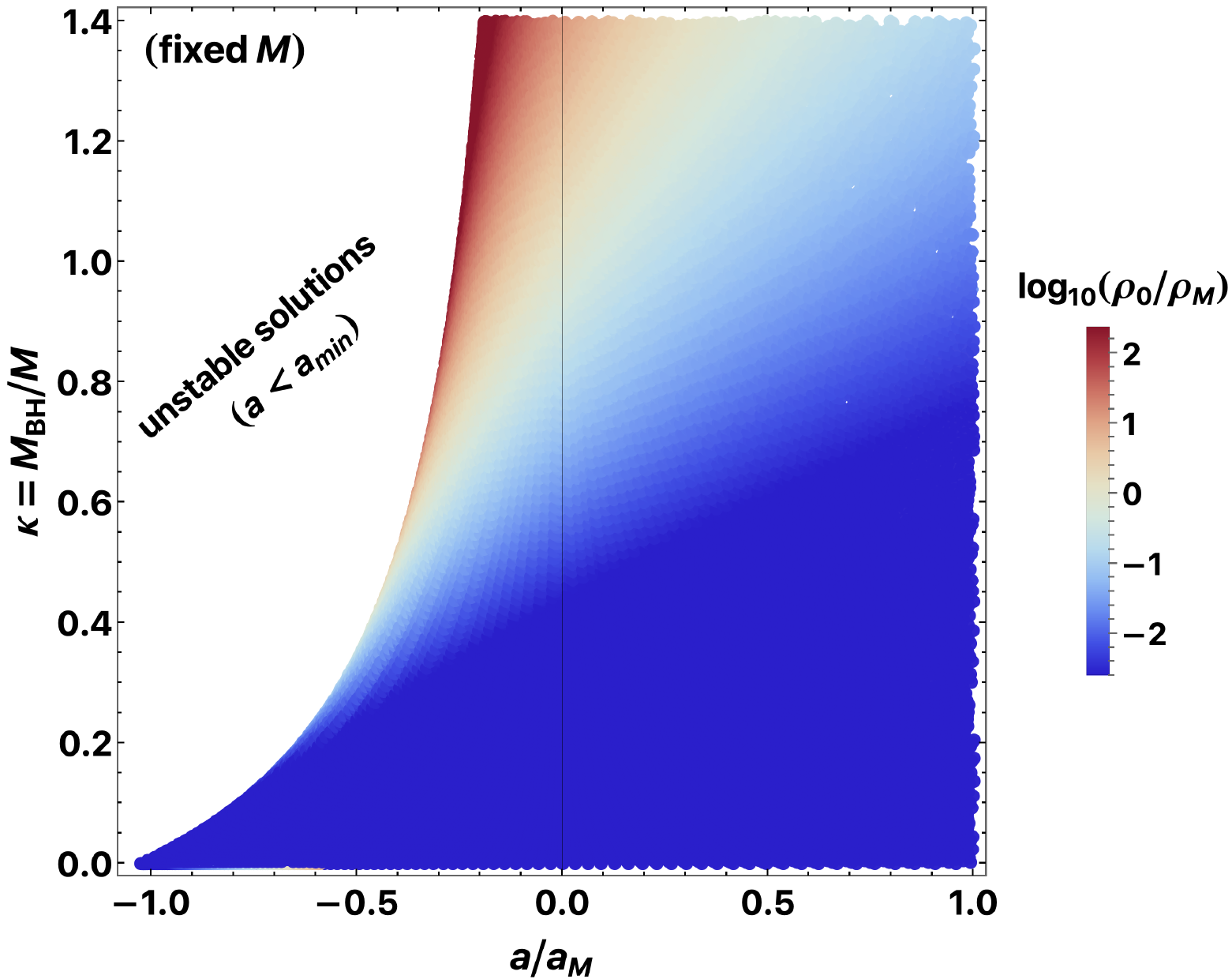}
    \\[2mm]
    \includegraphics[width=0.47\textwidth]{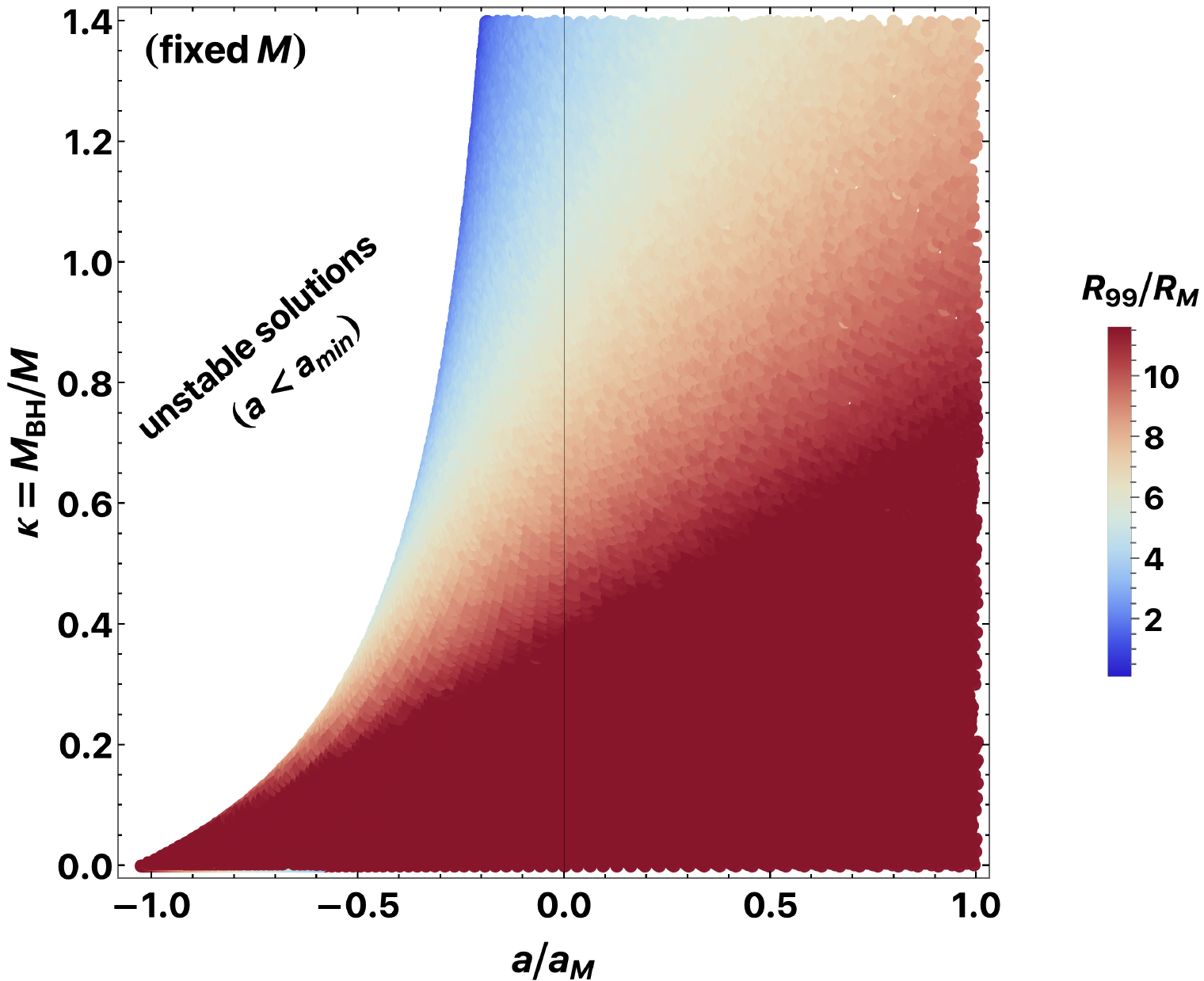}
    \caption{Allowed parameter space in the plane of the scattering length, normalized to $a_M \equiv (GM^2m)^{-1}$, for a boson star with self-couplings, of fixed mass $M$, hosting a central black hole. There exists a minimum scattering length $a_{\rm min}$, below which the solutions are unstable. Increasing the black hole mass increases this $a_{\rm min}$. \textit{Top}: Central density normalized to units of $\rho_M \equiv G^3 M^4m^6$. \textit{Bottom}: The radius containing $99\%$ of the boson star mass, normalized to units of $R_M\equiv (GM\,m^2)^{-1}$.}
    \label{fig:Mfix}
\end{figure}

We show the parametric dependence of $\rho_0$ and $R_{99}$ in the plane of the scattering length and the ratio between the masses of the black hole and the boson star in Fig.~\ref{fig:Mfix}. In this case, unstable solutions arise for attractive interactions with strengths below a minimum allowed value of the scattering length, $a_{\rm min}$. This minimum scattering length corresponds to the critical value $\lambda_*$ below which the quantum pressure can no longer balance the gravitational and self-interaction pressures, leading to the collapse of the system. Furthermore, this minimum scattering length depends on the ratio $\kappa$, and for fixed $M$ implies that increasing $M_{\rm BH}$, increases $a_{\rm min}$. This is due to the additional attractive gravitational force exerted by the black hole, which must be compensated by reducing the strength of attractive self-interactions. The densest and smallest boson stars are therefore obtained close to the region of unstable solutions. For repulsive interactions, there is no limitation on the strength $|a|$; however, for $a > 0$, the boson star is considerably less dense and more spread out.

\section{Analytic Approximations}
\label{sec:ansatz}
In the previous Section, we have demonstrated the procedure to obtain the density profile for a boson star hosting a black hole numerically by solving the differential equation \eqref{eq:red_hydro2} and appropriately scaling the solution back into physical coordinates. However, analytic insight can be gained by assuming an approximate solution for the density profile. To this end, we consider the following \textit{ansatz} for the density profile:
\begin{equation}
    \rho(r) = A\,e^{-r^2/R^2-2\beta r/R} \,,
    \label{eq:ansatz}
\end{equation}
where
\begin{equation}
    A \equiv \frac{M}{\pi^{3/2} R^3} \left[(1+2\beta^2)e^{\beta^2}\,{\rm erfc}(\beta)-\frac{2\beta}{\sqrt{\pi}}\right]^{-1}\,,
\end{equation}
arising from the relation between $\rho(r)$ and $M$ in \eqref{eq:mass_sol}. Here, ${\rm erfc}(x) \equiv 1 - {\rm erf}(x)$, with ${\rm erf(x)}$ being the error function. Following a similar approach in deriving \eqref{eq:n1}, we obtain
\begin{equation}
    \beta = G M_{\rm BH}m^2\,R\,.
\end{equation}

This approximate solution proposed is compatible with the expansion of the density profile around the center given in \eqref{eq:expansion}. The length scale $R$ specifies the characteristic radius of the boson star, depending primarily on the model parameters: the DM mass $m$, the self-interaction strength $\lambda$, and the total mass of the boson star, $M$. However, the introduction of the black hole induces an additional dependence of $R$ on the black hole mass, besides the usual dependence on the boson star parameters. The dimensionless parameter $\beta$ therefore captures the effects of the black hole within the boson star. Thus, \eqref{eq:ansatz} incorporates the dependence on the parameters $(m,\lambda,M,M_{\rm BH})$, consistent with the numerical solution to \eqref{eq:red_hydro2}. We note for completeness that for $M_{\rm BH} = 0$, we restore the results of a ``pure'' boson star, as studied in Refs.~\cite{Chavanis:2011zi,Chavanis:2011zm}.

Using this ansatz, we will explore various properties of the BS-BH system, such as the mass-radius relation and the dependence of the characteristic radius on the scattering length, comparing to our numerical results. To do so, we will follow the approach in Ref.~\cite{Chavanis:2011zi}, by first considering the total energy derived from \eqref{eq:action_non-rel}:
\begin{equation}
    \begin{aligned}
        E &= \int d^3r\,\left\{\frac{|\nabla \psi|^2}{2m} + m|\psi|^2\Phi+\frac{2\pi a|\psi|^4}{m}\right\}\, \\
          & \equiv \Theta + W + U\,,
    \end{aligned}
    \label{eq:energy_func}
\end{equation}
with the contributions stemming from the kinetic energy $\Theta$, the internal energy $U$ arising due to the self-interaction, and the gravitational potential energy $W$. We then insert the Madelung transformation \eqref{eq:madelung_transform} into \eqref{eq:energy_func}. The kinetic energy can be split into two components $\Theta = \Theta_C + \Theta_Q$, where
\begin{align}
    \Theta_C &= \int d^3r\,\rho\,\frac{|\vec{u}|^2}{2}\,,\\
    \Theta_Q &= -\frac{1}{m}\int d^3r\,\sqrt{\rho}\,\left(\frac{\nabla^2\sqrt{\rho}}{2m\sqrt{\rho}}\right)\,.
\end{align}
Similarly, the gravitational potential is sourced by the boson star and the black hole,
\begin{align}
    W = \int d^3r \,\rho\,\Phi = W_{\rm BS} + W_{\rm BH}\,,
\end{align}
where $\Phi$ is given by \eqref{eq:poisson_sol}. Finally, the internal energy is given by
\begin{equation}
    U = \frac{2\pi a}{m^3}\int d^3r \,\rho^2 \,.
\end{equation}
As we consider the time-independent solutions in hydrostatic equilibrium, where $\vec{u} = 0$, we may focus on the potential $V$ formed from quantum kinetic energy, the gravitational potential energy and the internal energy,
\begin{equation}
    V = \Theta_Q + W + U\,.
    \label{eq:potential}
\end{equation}
With the spherically symmetric profile considered in \eqref{eq:ansatz}, these have the explicit expressions,
\begin{subequations}
    \begin{align}
 &\begin{aligned}
     \Theta_Q &= \frac{1}{8m^2}\,\left[4\pi\int_0^{\infty}dr\, r^2\, \frac{1}{\rho(r)}\left(\frac{\partial \rho(r)}{\partial r}\right)^2\right] \\
              &= \sigma(\beta)\frac{M}{m^2 R^2} \,,
 \end{aligned}\label{eq:quantum_pot}
 \\
 &\begin{aligned}
     U &= \frac{2\pi a}{m^3}\left[4\pi\int_0^{\infty}dr\, r^2\,\rho^2(r)\right] \\
       &= \xi(\beta)\frac{2\pi a M^2}{m^3 R^3}\,,
 \end{aligned}\label{eq:internal_energy}
 \\
 &\begin{aligned}
     W_{\rm BS} &= -G\left[4\pi\int_0^{\infty}dr\, r\,\rho(r)\,M(r)\right] \\
                &= -\nu_1(\beta) \frac{G M^2}{R}\,,
 \end{aligned}\label{eq:grav_energy_BS}
 \\
 &\begin{aligned}
     W_{\rm BH} &= -G M_{\rm BH}\left[4\pi\int_0^{\infty}dr\, r\,\rho(r)\right] \\
                &= - \nu_2(\beta) \frac{G M M_{\rm BH}}{R}\,.
 \end{aligned}\label{eq:grav_energy_BH}
    \end{align}
\end{subequations}
Detailed expressions for the functions $\sigma, \xi, \nu_1$ and $\nu_2$ are provided in Appendix~\ref{app:ansatz_analysis}. We note that these coefficients are now increasing functions of $\beta$, implying for fixed boson star parameters, increasing the black hole mass increases the contribution of the relevant quantity to the potential~\eqref{eq:potential}. Note that on setting $M_{\rm BH} = 0$, we have $\sigma(0) = 3/4,\,\xi(0) = (2\pi)^{-3/2},\,\nu_1(0) = (2\pi)^{-1/2}$, the usual results for the pure Gaussian ansatz as in Ref.~\cite{Chavanis:2011zi}. This also gives $\nu_2(0) = 2/\sqrt{\pi}$, but the contribution $W_{\rm BH}$ vanishes in the potential.

\begin{figure}[htbp]
    \centering
    \includegraphics[width=0.48\textwidth]{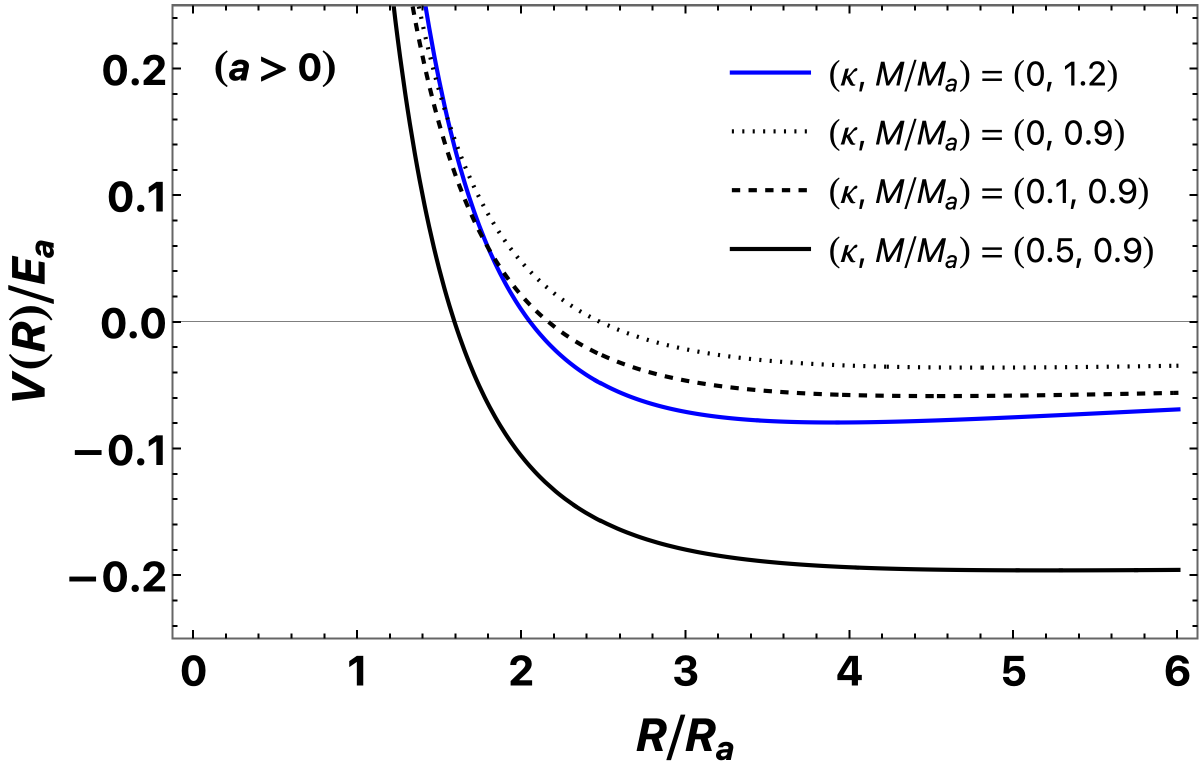}\\[2mm]
    \includegraphics[width=0.48\textwidth]{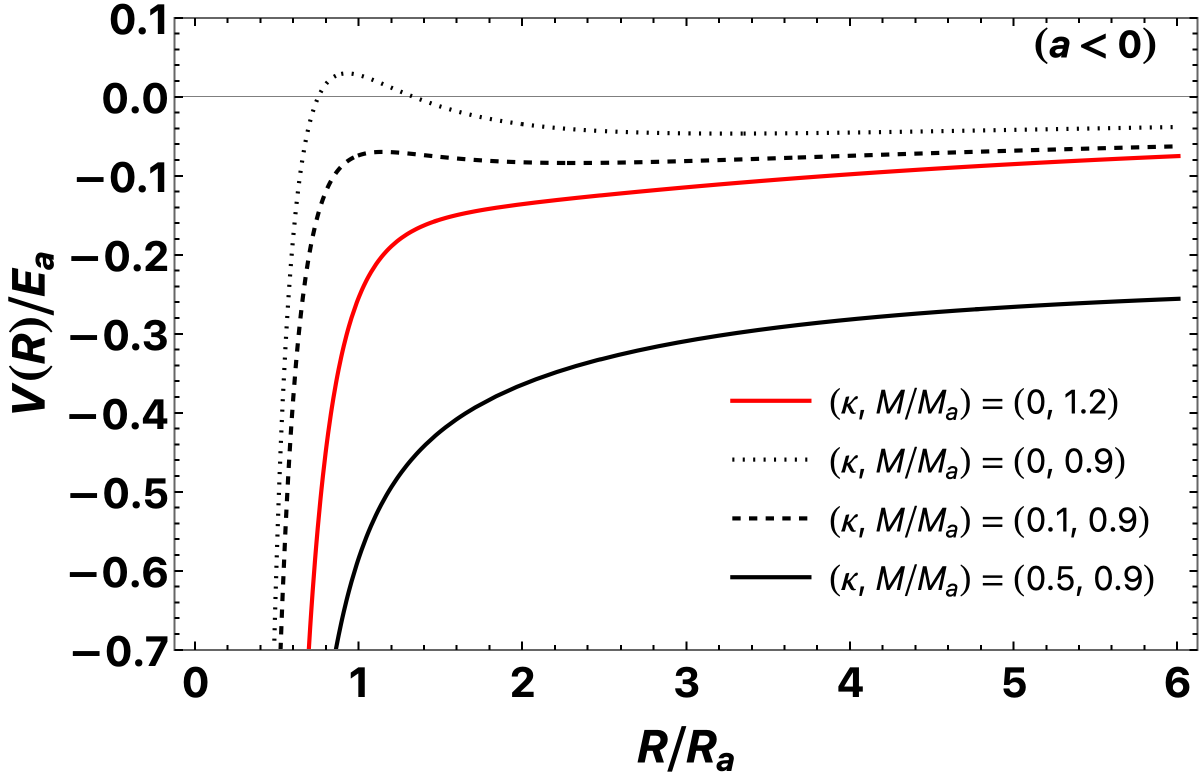}
    \caption{Potential in \eqref{eq:potential} as a function of the radius $R$ with all quantities entering rendered dimensionless through appropriate normalizations, $E_a \equiv Gm/R_a^2$ and $R_a \equiv \sqrt{|a|/(Gm^3)}$.
        \textit{Top}: For repulsive interactions, the potential is always bounded from below, with a single minimum, corresponding to a single stable configuration. Increasing $\kappa$ shifts the location of the minimum, but the system remains stable.
        \textit{Bottom}: For attractive interactions, without the central black hole, there exists a maximum mass $M_{\rm max}$ above which $V(R)$ is unbounded (solid red line), see main text for details. Below $M_{\rm max}$ (dotted black line), the potential exhibits a local maximum (corresponding to a smaller $R$) and a local minimum (corresponding to a larger $R$), which are associated with an unstable and a stable configuration, respectively. Introducing a black hole lowers the $M_{\rm max}$, and if $\kappa$ is large enough, i.e., for a large $M_{\rm BH}$, the potential is again unbounded.
    }
    \label{fig:potential_fixa}
\end{figure}

In Fig.~\ref{fig:potential_fixa}, we show the potential in \eqref{eq:potential} as a function of the radius $R$ for fixed magnitude of the scattering length. We adopt the normalization \eqref{eq:norm_fixa} to define the quantity $E_a \equiv Gm/R_a^2$, yielding the expression,
\begin{equation}
    \begin{aligned}
        \frac{V(\hat{R})}{E_a} &= \sigma(\beta)\frac{\hat{M}}{\hat{R}^2} + 2\pi \xi(\beta)\,{\rm sgn}(a)\frac{\hat{M}}{\hat{R}^3} \\
                               &\quad - \frac{\left(\nu_1(\beta) - \nu_2(\beta)\kappa\right)\,\hat{M}^2}{\hat{R}}\,,
    \end{aligned}
\end{equation}
where we have defined $\hat{M} \equiv M/M_a$ and $\hat{R} \equiv R/R_a$ for brevity. For repulsive interactions (positive sign of $a$), we observe that $V(\hat{R})$ is always bounded from below for various values of $\hat{M}$ and there is an equilibrium point, which is the global minimum. Despite increasing $\kappa$, implying increasing the gravitational potential of the black hole, the potential remains bounded from below, though the depth of the minimum increases. In contrast, for attractive interactions (negative sign of $a$) for particular values of $\hat{M}$, there can exist a maximum and a minimum of the potential. For $\hat{R}$ below the radius corresponding to the maximum, the potential becomes unbounded. If $\hat{M}$ is increased further, the potential can become unbounded, implying unstable solutions. This corresponds to the quantum potential being unable to balance the gravitational and interaction potentials, thereby setting a limit on the maximum mass that can be allowed for stability. Furthermore, if one considers a situation where the $\hat{M}$ corresponding to a stable solution is fixed, and now increases $\kappa$, the potential once again can become unbounded. These observations are consistent qualitatively with our discussion in Sec.~\ref{subsec:fix_a}.

\begin{figure}[htbp]
    \centering
    \includegraphics[width=0.48\textwidth]{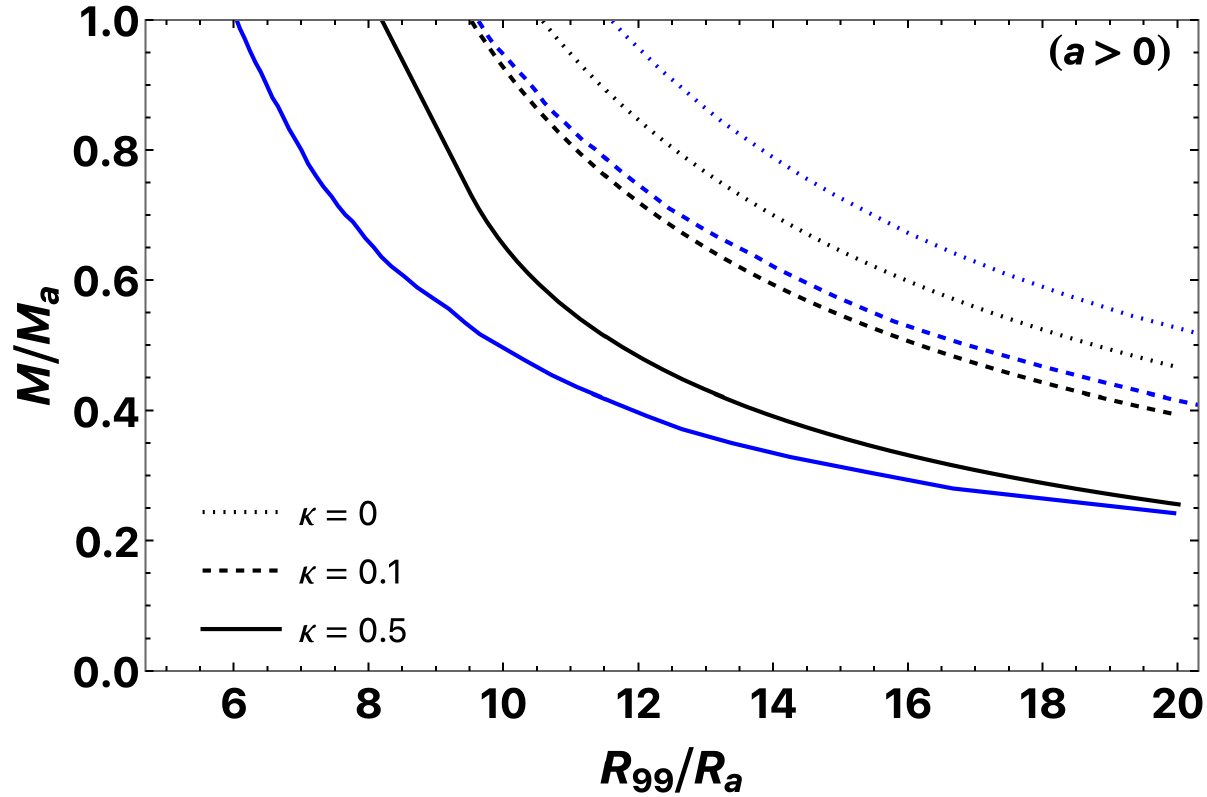}\\[2mm]
    \includegraphics[width=0.48\textwidth]{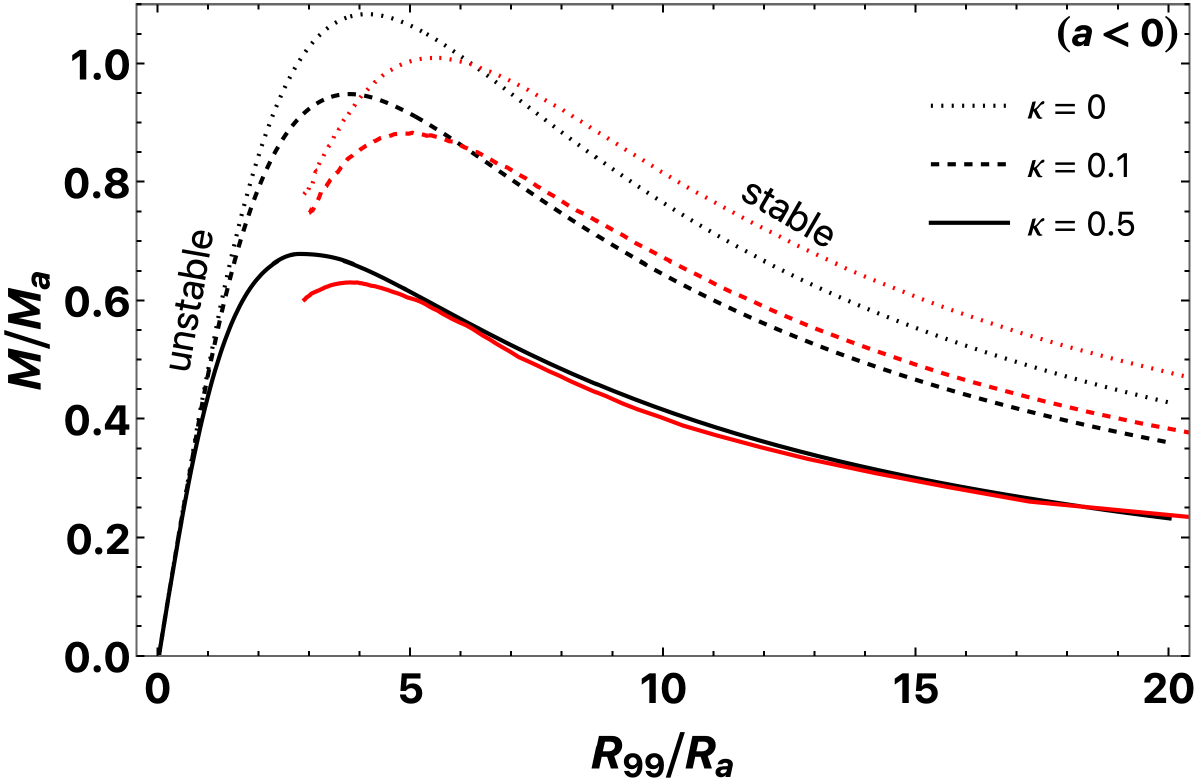}
    \caption{Mass-radius relation of the BS-BH system, for repulsive (\textit{top}) and attractive (\textit{bottom}) interactions, as given in Eq.~\eqref{eq:mass-rad_rel}, for various $\kappa$. We have normalized the mass of the boson star by $M_a \equiv (Gm|a|)^{-1/2}$ and $R_{99}$ by $R_a \equiv \sqrt{|a|/(Gm^3)}$. Colored lines are obtained from our numerical method. For attractive interactions, we have indicated the regions of unstable and stable configurations, lying to the left and right of a critical radius $R_{99}^*$, respectively, for the same mass $M < M_{\rm max}$. The unstable branch appears incomplete for the numerical method due to the range of input parameters we have considered to solve the differential equation \eqref{eq:red_hydro2}.
    }
    \label{fig:ansatz_afix_att_rep}
\end{figure}

We now seek the mass-radius relation of this system. Accordingly, the approach is to minimize \eqref{eq:potential} with respect to $R$, for given masses $M$ and $M_{\rm BH}$, and obtain the equilibrium radius. To this end, we can then inspect these minima through
\begin{equation}
    \frac{\partial V}{\partial R} = 0\,.
    \label{eq:crit_points}
\end{equation}
This leads to the following expression:
\begin{align}
    M(R)
 &= \frac{2\sigma}{\nu_1\,Gm^2R}\left[\left(1-\frac{1}{2}\frac{d\ln \sigma}{d \ln \beta}\right)-\frac{\beta\,\nu_2}{\nu_1}\left(1-\frac{d\ln\nu_2}{d\ln\beta}\right)\right]\nonumber \\
 &\quad\times \bigg[\left(1-\frac{d\ln\nu_1}{d\ln\beta}\right) -\frac{ 6\pi a}{Gm^3R^2}\frac{\xi}{\nu_1}\left(1-\frac{1}{3}\frac{d\ln\xi}{d\ln\beta}\right)\bigg]^{-1}\,,
 \label{eq:mass-rad_rel}
\end{align}
where we have suppressed the $\beta$ dependence of the various functions for brevity. We show the relation \eqref{eq:mass-rad_rel} for different repulsive and attractive interactions in Fig.~\ref{fig:ansatz_afix_att_rep}. To facilitate comparison with our numeric results, we normalize to the corresponding $R_{99}$ obtained from our ansatz \eqref{eq:ansatz}.
This quantity also depends on the mass of black hole through $\beta$, as shown in Fig.~\ref{fig:R99} of Appendix~\ref{app:ansatz_analysis}.
In particular, $R_{99}$ decreases with increasing $\beta$, i.e., larger $M_{\rm BH}$, from the value $\sim 2.382R$ for $\beta = 0$.

For $a > 0$, as already mentioned, due to the presence of a single extremum of the potential, there exists only one particular $M$ corresponding to the radius $R$ and fixed black hole mass. The mass is always a decreasing function of the characteristic radius. Furthermore, on increasing $\kappa$, implying increasing the black hole mass, the mass at a fixed $R$ decreases. This is due to the gravitational force of the black hole compacting the star. On comparing to our numerical results, we find the qualitative agreement in the trends, with a deviation at most a few percent for the mass-radius relation for $\kappa \lesssim 0.3$, with larger deviations occurring for larger $\kappa$.

For $a < 0$, we find that there exists a maximum mass $M_{\rm max}$ occurring at a particular radius $R_{99}^*$, for a fixed black hole mass. Above $M_{\rm max}$, there is no solution to \eqref{eq:crit_points}, as there are no equilibrium points for the potential, cf. Fig.~\ref{fig:potential_fixa}. For any $M < M_{\rm max}$, there exist two possible radii corresponding to the same $M$. The stable configuration has this value of $M$ at the larger radius, which is a local minimum of the potential \eqref{eq:potential}. The unstable configuration corresponds to the local maximum and occurs at a smaller radius, as shown in Fig.~\ref{fig:ansatz_afix_att_rep}. Within the stable solutions, similar to repulsive interactions, $M$ decreases as a function of $R$. Next, increasing the black hole mass has the effect of shrinking the boson star, i.e., for the same $M$, $R$ is smaller for increased $\kappa$. These observations are consistent with the inferences of Sec.~\ref{subsec:fix_a}. On further comparison with our numerical results, besides qualitative agreement, we find good quantitative agreement of the mass-radius relation obtained using our ansatz up to a percent level. For completeness, we note that the unstable branch from our numerical method is not fully present due to the range of the control parameters $(f_1, \bar{\chi})$ we have considered to solve \eqref{eq:red_hydro2}.

\begin{figure}[htbp]
    \centering
    \includegraphics[width=0.48\textwidth]{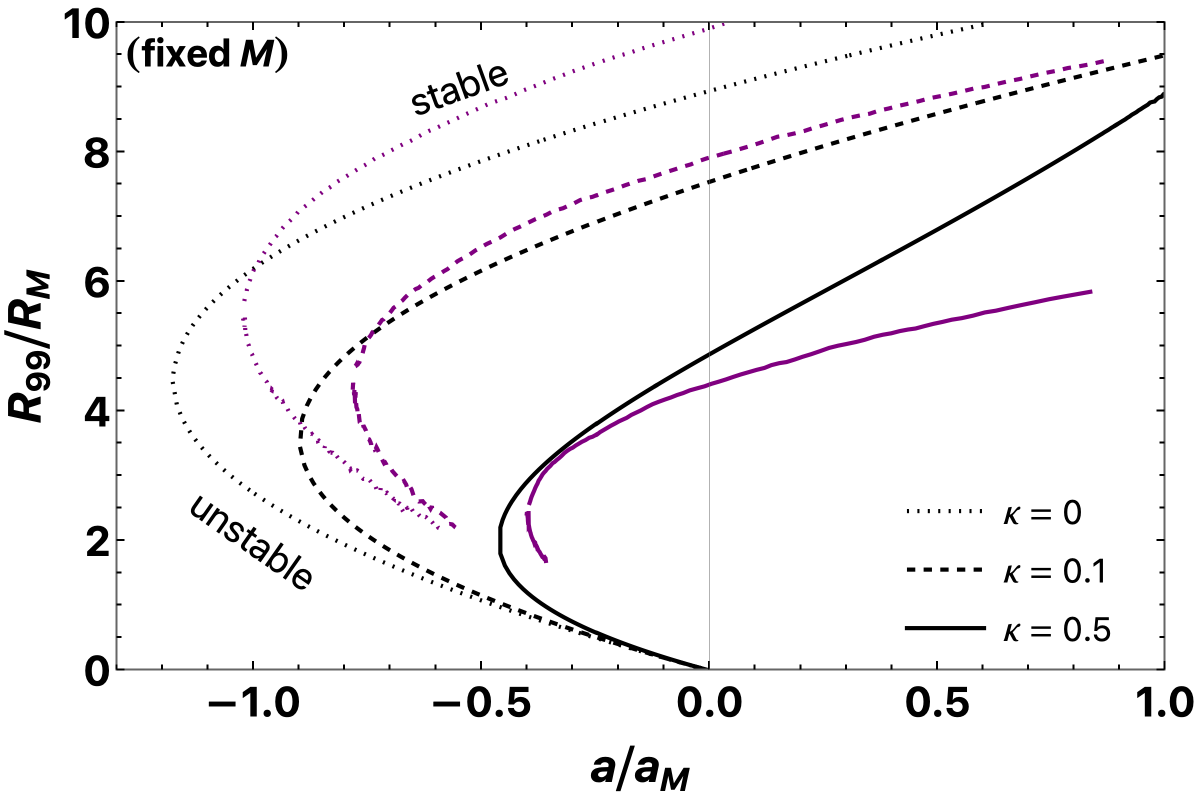}
    \caption{Radius containing $99\%$ of the boson star mass, normalized to $R_M\equiv (GM\,m^2)^{-1}$, as a function of the scattering length, normalized by $a_M \equiv (GM^2m)^{-1}$, for fixed boson star mass and various values of $\kappa$. Black lines are obtained by inverting \eqref{eq:mass-rad_rel} for fixed $M$. Purple lines are obtained from the exact numerical solution. For attractive interactions, the branch where $R_{99}$ increases for increasing $|a|$ is unstable. The unstable branches from our numerical method appear incomplete due to the finite window of input parameters for which we solve the differential equation \eqref{eq:red_hydro2}.}
    \label{fig:ansatz_Mfix}
\end{figure}

Additionally, from \eqref{eq:mass-rad_rel}, by (numerically) inverting the relation, we can obtain the radius as a function of the boson star parameters $(m, \lambda, M)$ and the black hole mass. Having already discussed the mass-radius relation, we now fix the mass of the boson star, and consider the radius as a function of the scattering length, which we show in Fig.~\ref{fig:ansatz_Mfix}. We adopt the normalization \eqref{eq:norm_fixM}, and once again scale the radius by $R_{99}$ to facilitate comparison with the numerical results.

For repulsive and no interactions $a \geq 0$, the radius is a monotonically increasing function of the scattering length, implying that for the same mass $M$, the boson star with larger $a$ will be more spread out. For attractive interactions, there is a minimum scattering length $a_{\rm min}$, below which no solutions are found. For $a > a_{\rm min}$, there exist two possible values of the radius, the stable solution being the one with the larger radius. This is again due to the requirement of minimizing the energy of the system. On introducing a black hole at the center, we see that the radius corresponding to the same $a$ decreases, thereby shrinking the boson star, due to the gravitational pull of the black hole. Correspondingly, the minimum allowed scattering length now increases when $\kappa$ is increased to permit a stable configuration of the system. This behavior is in line with the inferences of Sec.~\ref{subsec:fix_M}. Finally, we find that the curves in Fig.~\ref{fig:ansatz_Mfix} obtained from the ansatz agree well with the numeric results, with deviations of up to $10\%$, mainly for repulsive interactions and for larger $\kappa$. The incompleteness of the unstable branch from our numerical method is again associated with the finite range of the control parameters $(f_1, \bar{\chi})$, in solving \eqref{eq:red_hydro2}.

\section{Gravitational Wave Probes}
\label{sec:gw_probe}
In the previous Sections, we explored the properties of configurations of a boson star hosting a central black hole. We now consider the scenario where a second, smaller black hole of mass $M_2$ is captured by a BS-BH system, whose central black hole has mass $M_1$, such that $q \equiv M_2/M_1 \ll 1 $. The resulting binary system can emit gravitational waves during the inspiral phase, and as we show, the GW waveform is modified on account of the boson star environment. Therefore, the GWs from such a binary system, if detected by LISA, may offer insight into the properties of the BS-BH system.

\subsection{Binary evolution}

Given the surrounding boson star density around the central black hole, the equation of motion in the radial direction is given by,
\begin{equation}
    -\ddot{r} + r\,\omega_s^2 = G\frac{M_{\rm tot}}{r^2} \,,
\end{equation}
where $r$ is the relative distance between the binary constituents, $\omega_s$ is the angular velocity and the total mass of the binary system is given by $M_{\rm tot} \equiv \tilde{M}_1 + M_2$, with $\tilde{M}_1$ being the total mass enclosed in the orbit,
\begin{equation}
    \tilde{M}_1 = \begin{cases}
        M_1, & r < r_{\rm ISCO} \,, \\
        M_1 + 4\pi\int_{r_{\rm ISCO}}^{r}du\,u^2\,\rho(u) \,, & r \geq r_{\rm ISCO}\,,
    \end{cases}
    \label{eq:mass_enc}
\end{equation}
where $r_{\rm ISCO} = 3 r_s$ is defined as the innermost stable circular orbit (ISCO), and the radius at which we take the merger to complete, with $r_s = 2 G M_1/c^2$ being the Schwarzschild radius of the central black hole, and $c$ being the speed of light in vacuum. Equation~\eqref{eq:mass_enc} incorporates the additional mass due to the surrounding boson star environment around the central black hole. Since we will be interested in the cases where $q \ll 1$, the orbital motion can be treated as quasicircular, with the boson star environment considered as nearly unperturbed~\cite{Kadota:2023wlm, Banik:2025fnc}. In this approximation, $\ddot{r}$ vanishes and we have the usual Keplerian relation for the angular velocity,
\begin{equation}
    \omega_s = \sqrt{\frac{G M_{\rm tot}}{r^3}}\,.
    \label{eq:orb_freq}
\end{equation}

Besides the gravitational pull of DM inside the orbit, the host boson star environment can affect the evolution of the binary system by imparting DF~\cite{Chandrasekhar:1943ys}, thus accelerating the merger rate of the binary in comparison to pure GW emission.
From energy conservation, the energy lost in the orbital motion is due to GW emission and dynamical friction,
\begin{equation}
    -\dot{E}_{\rm orb} = P_{\rm GW} + P_{\rm DF}\,,
\end{equation}
where $P_{..}=v\,F_{..}$, with $v = r\omega_s$ being the velocity of the smaller black hole with mass $M_2$.
This gives the evolution of the distance between the two BHs as
\begin{equation}
    \mu\,\dot{r} = -\left(F_{\rm GW} + F_{\rm DF}\right)\left(2\omega_s +r\frac{d\omega_s}{dr}\right)^{-1}\,,
\end{equation}
where $\mu \equiv \tilde{M}_1\,M_2/M_{\rm tot}$ is the reduced mass of the system.
The force resulting in GW emission is given by~\cite{Maggiore:2007ulw}
\begin{equation}
    F_{\rm GW} = \frac{1}{v}\,\frac{32G^4\mu^2 M_{\rm tot}^3}{5c^5r^5}\,.
\end{equation}

\begin{figure}[htbp]
    \centering
    \includegraphics[width=0.48\textwidth]{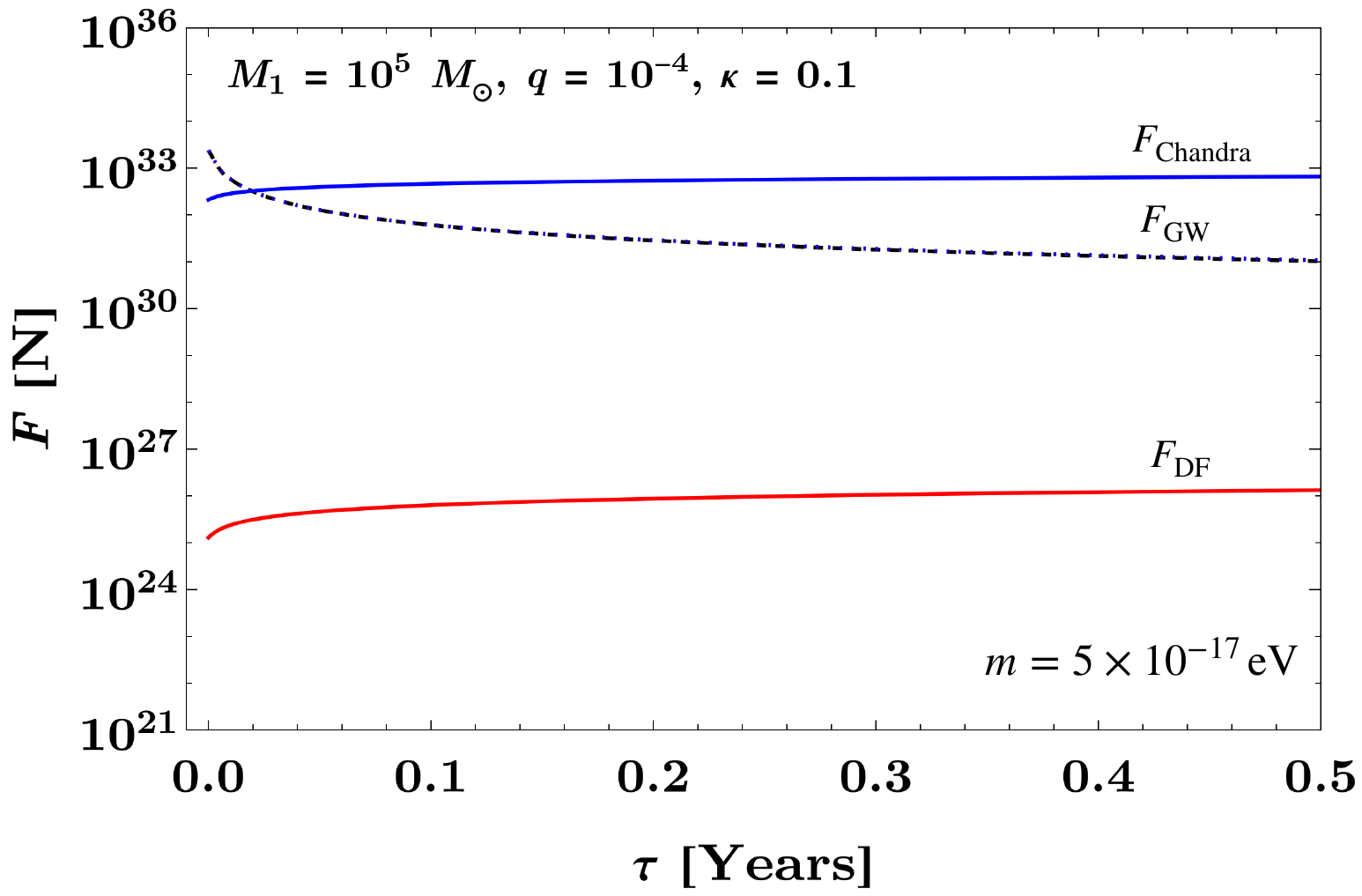}
    \caption{Evolution of the various forces resulting in energy loss over the observation time of $0.5$~years, with $\tau$ representing the time to reach ISCO.
        Here, we have $\lambda = 0$, corresponding to a self-gravitating BS-BH system with a secondary black hole inspiralling around it. We compare the force of dynamical friction based on Chandrasekhar's formula in \eqref{eq:chandra_DF} with the formula taking into account the wavelike nature of ULDM, cf. \eqref{eq:ULDM_DF}. The force emitted from GW is shown for comparison, which is subdominant when compared to $F_{\rm Chandra}$, but is dominant when using $F_{\rm DF}$.
    }
    \label{fig:FDF_comparison}
\end{figure}

The DF force experienced by the secondary black hole inside a DM environment is often estimated using Chandrasekhar's formula~\cite{Chandrasekhar:1943ys},
\begin{equation}
    F_{\rm Chandra} = 4\pi\frac{(GM_2)^2\,\rho(r)}{v^2}\,\ln\Lambda
    \label{eq:chandra_DF}
\end{equation}
where $\ln \Lambda = \ln \sqrt{M_1/M_2}$ is the widely used approximation in the literature~\cite{2008gady.book.....B, 2010gfe..book.....M,Kavanagh:2020cfn} for the Coulomb logarithm. However, for ULDM, the quantum pressure can affect the accumulation of DM particles in the wake behind the smaller BH~\cite{Lancaster:2019mde}. To account for this effect, we adopt the following formula:\footnote{Although the boson star environment is an ULDM background with self-interactions, we neglect these in the computation of the dynamical friction, effectively treating the background as ``fuzzy DM''. For the parameter space we naturally arrive in, as discussed at the end of this section, this is a reasonable approximation as the quantum pressure dominates over the pressure exerted by the self-interactions.
}
\begin{equation}
    F_{\rm DF} = 4\pi\frac{(GM_2)^2\,\rho(r)}{v^2} C_{\rm rel} ,
    \label{eq:ULDM_DF}
\end{equation}
where $C_{\rm rel}$ is a function of $\Lambda$ and the ``quantum Mach number'' $\mathcal{M}_Q$ , which is defined as
\begin{equation}
    \mathcal{M}_Q \equiv \frac{v}{v_Q}\,,
    \label{eq:quantum_mach}
\end{equation}
i.e., the ratio between the velocity of the secondary black hole and $v_Q \equiv GMm$.
In the limit $\mathcal{M}_Q \to 0$, which occurs for larger DM masses, $C_{\rm rel}$ approaches $\ln\Lambda$, therefore recovering the result \eqref{eq:chandra_DF}. However, in the cases of interest in our analysis, we have $\mathcal{M}_Q \gg 1$, for which $C_{\rm rel}$ is given by~\cite{Lancaster:2019mde}
\begin{equation}
    C_{\rm rel} = {\rm Cin}\left({2\Lambda}/{\mathcal{M}_Q}\right)+\frac{\sin (2 \Lambda/\mathcal{M}_Q)}{2 \Lambda/\mathcal{M}_Q}-1\,.
    \label{eq:Crel}
\end{equation}
Here, ${\rm Cin}(x) \equiv \int_0^x dt (1-\cos t)/t $ is the cosine integral. This coefficient $C_{\rm rel}$ can result in a reduction of up to seven orders of magnitude in the DF compared to Chandrasekhar's formula, for $m = 5\times 10^{-17}\,{\rm eV}$, as shown in Fig.~\ref{fig:FDF_comparison}. In the same figure, we observe that GW emission is the subdominant force driving the energy loss in the system when compared to the DF as predicted by Chandrashekhar's formula, due to the high density of the BS-BH system. Also, $F_{\rm GW}$ increases as the binary approaches merger, due to the $\sim r^{-9/2}$ dependence. In contrast, when one uses Eq.~\eqref{eq:ULDM_DF}, which properly takes into account the accumulation of ULDM in the wake of the secondary black hole, $F_{\rm GW}$ is the dominant driving force. We also note that both predictions of the DF force decrease slightly toward the end of the merger. This can be attributed to the increase in the velocity of the secondary black hole because of the $v \sim r^{-1/2}$ dependence being dominant over the increase of the density of the BS-BH system throughout the inspiral range considered (see Fig.~\ref{fig:density_profile_eg}). Thus, the numerators of \eqref{eq:chandra_DF} and \eqref{eq:ULDM_DF} increase more slowly than their corresponding denominators, resulting in an overall decrease in the dynamical friction force.

\subsection{Gravitational wave analysis}

The GW waveforms emitted from the inspiral of the binary system are given by~\cite{Maggiore:2007ulw}
\begin{subequations}
    \begin{align}
        h_{+}(t) =
        &\frac{4}{D_L} \left(\frac{GM_c}{c^2}\right)^{5/3}\,\left[\frac{\pi f(t_{\rm ret})}{c}\right]^{2/3}\,\frac{(1+\cos^2\iota)}{2}\nonumber\\
        &\times\cos\left[\Psi(t_{\rm ret})\right]\,,
        \\
        h_{\times}(t) =
        &\frac{4}{D_L} \left(\frac{GM_c}{c^2}\right)^{5/3}\,\left[\frac{\pi f(t_{\rm ret})}{c}\right]^{2/3}\,\cos\iota\nonumber\\
        &\times\sin\left[\Psi(t_{\rm ret})\right]\,,
    \end{align}
\end{subequations}
where $D_L$ is the luminosity distance to the binary source, $M_c = \mu^{3/5} M_{\rm tot}^{2/5}$ is the chirp mass, $t_{\rm ret} \equiv t - D_L/c$ is the retarded time, $\iota$ is is the angle between the orbital angular momentum axis of the binary and the detector direction, and $f$ is the GW frequency, related to \eqref{eq:orb_freq} as $f \equiv 2\omega_s/(2\pi)$.

\begin{figure}[htbp]
    \centering
    \includegraphics[width=0.48\textwidth]{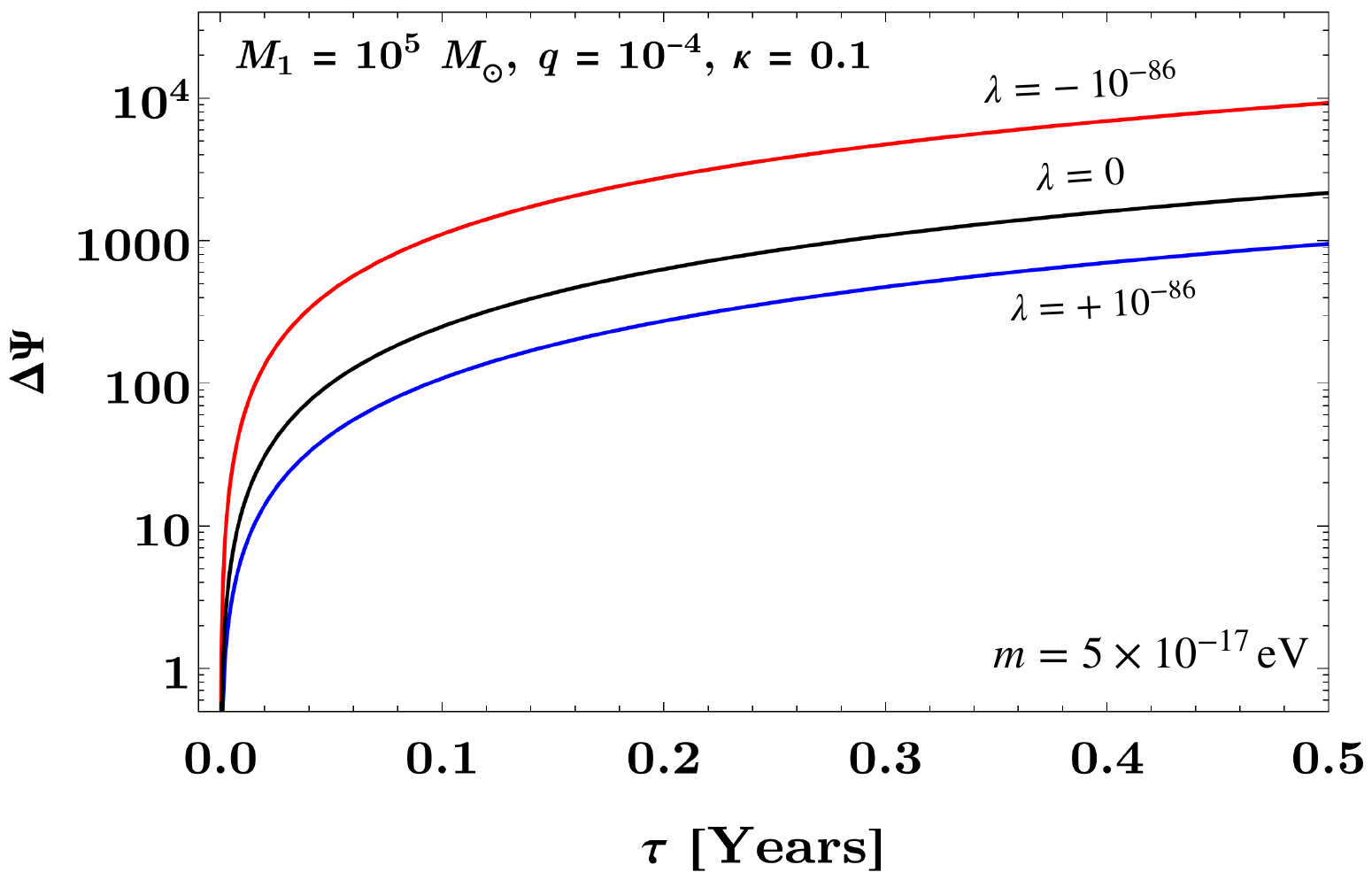}
    \caption{Dephasing effect for inspirals of duration 0.5~years for different interaction types of the BS-BH system with a secondary black hole. The GWs emitted from inspirals within a boson star with attractive interactions experience the most dephasing due to the larger central density.
    }
    \label{fig:dephasing}
\end{figure}

As mentioned, in comparison to a binary inspiralling in vacuum (i.e., without any boson star environment), the binary inspiral within the environment of the boson star experiences dynamical friction, which accelerates the merger rate. This reduces the number of orbital cycles to merger, given by,
\begin{equation}
    N_{\rm{cyc}} \equiv \int_{0}^{t_{\rm obs}} f(\tau') \,d\tau',
\end{equation}
where $t_{\rm obs}$ is the observation time for the merger to occur. One associates this to the phase of the GWs, such that $\Psi = 2\pi N_{\rm cyc}$, and therefore the \textit{dephasing} is defined as,
\begin{equation}
    \Delta \Psi = \Psi_{\rm{vac}} - \Psi_{\rm{BS}}\,.
    \label{eq:dephase}
\end{equation}
This effect is demonstrated in Fig.~\ref{fig:dephasing}, assuming an observation time of half a year, for boson stars with different interaction types. As attractive interactions increase the overall central density of the boson star, this results in a larger dephasing in comparison to the self-gravitating case. Repulsive interactions show the least dephasing, as the boson star is now thinner and spread out.

We now leverage this dephasing effect to probe properties of boson stars. To this end, we consider the GW signal strain in LISA~\cite{Rubbo:2003ap},
\begin{align}
    h(t) &=  h_{+}(t-\Delta t)\,F_{+}(\vartheta,\varphi,\varsigma,t-\Delta t) \nonumber \\
         &\quad + h_{\times}(t-\Delta t)\,F_{+}(\vartheta,\varphi,\varsigma,t-\Delta t)\,
         \label{eq:GWstrain_det}
\end{align}
where we have chosen the polar coordinate system with the Sun at its origin. Here, $F_{+}$ and $F_{\times}$ are the detector response functions for LISA, which we provide in Appendix~\ref{app:LISA_detect}, which depend on the latitude $(\vartheta)$ and longitude $(\varphi)$ of the binary in the polar coordinate system, the polarization angle $(\varsigma)$ and the time arrival delay of the GWs between the Sun and the detector.

To quantify the detection prospects, one requires the signal-to-noise ratio (SNR) of the GW signal in the detector given by
\begin{equation}
    {\rm SNR} \equiv \sqrt{4\int_{f_{\rm min}}^{f_{\rm max}} |d(f)|^2 \,df}\quad {\rm where} \quad d(f) = \frac{\tilde{h}(f)}{\sqrt{S_n(f)}}\,.
\end{equation}
Here $\tilde{h}(f)$ is the Fourier transform of the time domain signal in \eqref{eq:GWstrain_det}, $S_n(f)$ noise power spectral density of the detector and $f_{\rm max}$ and $f_{\rm min}$ are the frequencies at the end and beginning of the observation, respectively. If the computed SNR is greater than a threshold (often ${\rm SNR}_{\rm thr} = 10$), then the signal can be detected by LISA.

Although the SNR informs us about the region of parameter space feasible for detection, it does not quantify how precisely the parameters determining the signal can be measured. To address this, we consider a Fisher forecast analysis in the region where the SNR is high, which we take as ${\rm SNR} = 100$. In this case, the posterior probability distribution of the parameter set determining the GW signal (given by $\theta_i$) can be approximated by a multivariate Gaussian distribution centered around the true values (corresponding to $\hat{\theta}$). The Fisher information matrix is defined as
\begin{equation}
    \Gamma_{ij} \equiv \left(\frac{\partial d(f)}{\partial \theta_i}, \frac{\partial d(f)}{\partial \theta_j} \right)\bigg|_{\theta = \hat{\theta}}\,,
\end{equation}
where we have defined the bracket operator
\begin{equation}
    (X,Y) = 2\int_{f_{\rm min}}^{f_{\rm max}} df \left[X(f)Y(f)^{\ast}+X(f)^{\ast}Y(f)\right]\,.
\end{equation}
From the inverse of the Fisher matrix $\Sigma \equiv \Gamma^{-1}$, we can obtain the $1\sigma$ errors of the parameters from the diagonal elements as
\begin{equation}
    \sigma_{\theta_i} = \sqrt{\Sigma_{ii}}\,,
\end{equation}
and the correlation coefficients,
\begin{equation}
    c_{\theta_i \theta_j} = \frac{\Sigma_{ij}}{\sigma_{\theta_i} \sigma_{\theta_j}}\,.
    \label{eq:corr_coeff}
\end{equation}

For the binary system, consisting of a central black hole hosted by a boson star and a secondary black hole, the parameter vector is given by
\begin{equation}
    \theta = \{M, m, \lambda;
    M_1,M_2,D_L,\iota,\varsigma,\vartheta,\varphi,\phi_{\rm ISCO},t_{\rm ISCO}\}\,.
\end{equation}
Of these, the first three parameters determine the boson star (i.e. DM) properties, and the rest relate to the binary system. We set the source phase $\phi_{\rm ISCO}$ and time $t_{\rm ISCO}$ at ISCO to zero, and vary the luminosity distance $D_L$ to ensure a high SNR. Furthermore, as our interest lies in studying the impact of varying the DM parameters on the signal, we fix the binary system with $(M_1,\,M_2) = (10^5\,M_{\odot}, 10\,M_{\odot})$, and the various angles $\{\iota,\varsigma,\vartheta,\varphi\}$ are set to $\pi/4$.

For a stable numeric implementation, we found it convenient to enter the boson star parameters in the form,
\begin{equation}
    \{M,m,\lambda\}\to \{\kappa,\rho_M, a/a_M\}\,.
\end{equation}
This is motivated as follows: given that $M_1$ (the mass of the central BH) is fixed, $\kappa$ determines the boson star mass as $M = M_1/\kappa$. Then, the quantity $\rho_M$, given in \eqref{eq:rhoM}, serves to set the overall density of the boson star and is a proxy for $m$, the mass of the constituent ULDM. Finally, enhancement or suppression to this density is determined by the strength and sign of the coupling $\lambda$, which we provide through $a/a_M$, whereby $a_M$ is fixed for a given $M$ and $m$.

\begin{figure}[t!]
    \centering
    \includegraphics[width=0.48\textwidth]{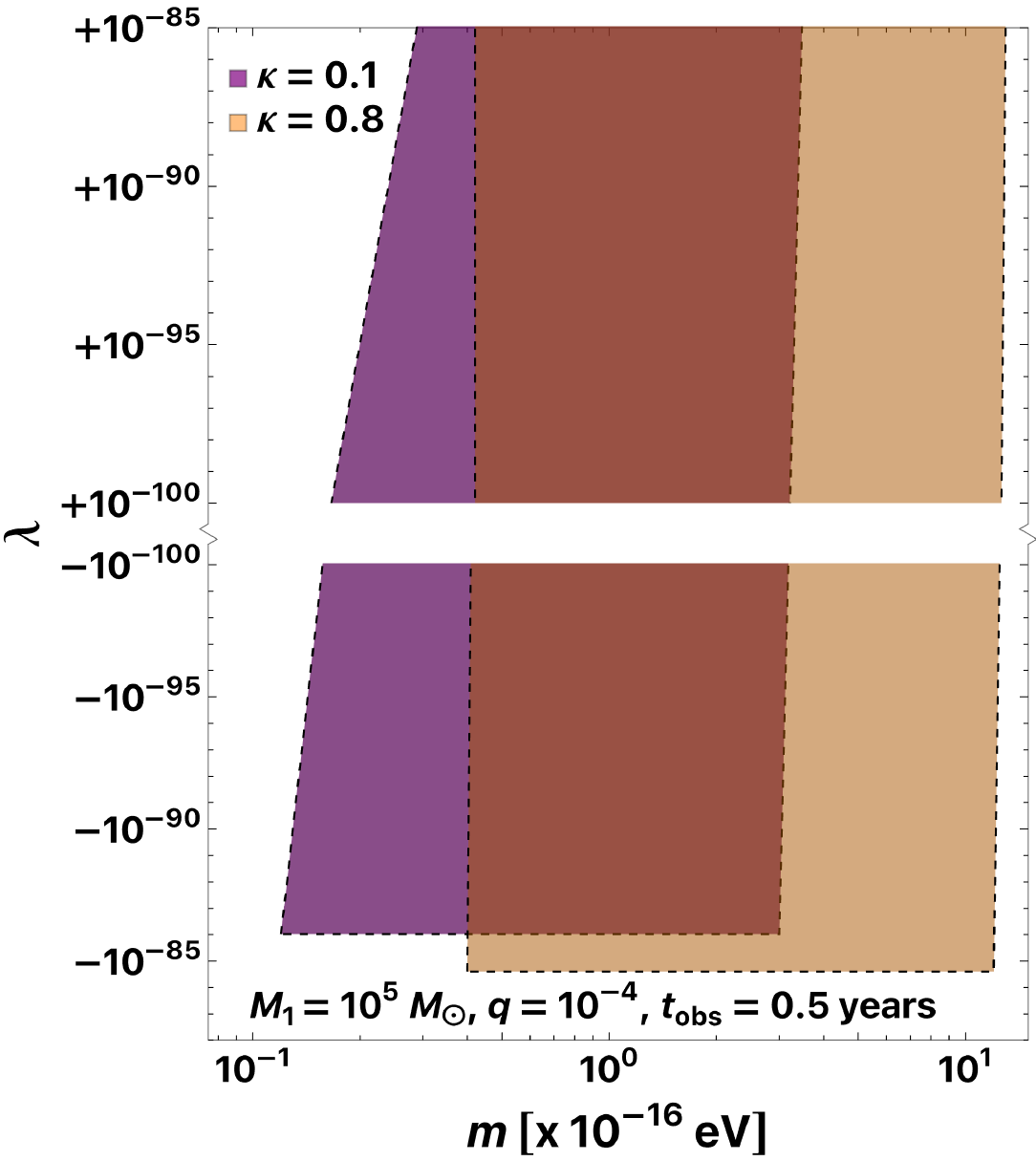}
    \caption{Regions of parameter space in the plane of the ULDM mass and the quartic coupling, where, based on the dephasing of GWs, the boson star parameters can be pinned down to $1\sigma$ uncertainty. We consider an inspiral observed for half a year, with the central black hole having mass $M_1 = 10^5\,M_{\odot}$ and the secondary black hole with mass $M_2 = 10\,M_{\odot}$. We show the scenarios with $\kappa = 0.1$ (purple) and $\kappa = 0.8$ (orange), across both attractive and repulsive interactions. The white space separating the positive and negative $\lambda$ regions is also detectable, corresponding to the limit of no interactions, marked with a kink on the $y$ axis. The detectable region begins for a minimum mass of ULDM to arrive at the minimum density for observable dephasing. For larger masses, the system becomes highly dense, but thinner, resulting in the halting of GW dephasing. The lower bound on negative $\lambda$ arises from the requirement of stable configurations of the BS-BH system. There is expected to exist an upper bound on $\lambda$ for repulsive interactions (not shown) above which the system becomes too sparse to allow sufficient GW dephasing.
    }
    \label{fig:constraints}
\end{figure}

Given half a year of observation time by LISA, we can examine the relative uncertainty on the measurement of the boson star parameters. For a fixed $\kappa$, by demanding the relative uncertainties on the $\rho_M$ and the $a/a_M$ are less than $10\%$, we can identify detectable regions corresponding to the parameter space of $(m,\lambda)$, as shown in Fig.~\ref{fig:constraints}. We find that to pin down these parameters to the given uncertainty limit, we require the central density to be $\rho \sim 10^{22}\,M_{\odot}{\rm pc}^{-3}$, which is higher than previous works~\cite{Kadota:2023wlm, Boudon:2023vzl}. This requirement arises from considering the formula \eqref{eq:ULDM_DF} for the DF force; in the aforementioned regions of parameter space for the ULDM mass we have considered, Chandrasekhar's formula in \eqref{eq:chandra_DF} overpredicts the DF force.

Given the central black hole mass $M_1 = 10^5\,M_{\odot}$, this implies that we have $m \sim 10^{-17}-10^{-15}\,{\rm eV}$ for the two choices of $\kappa$ considered. For $\kappa = 0.1$, giving $M = 10^6\,M_{\odot}$, we find that for a lower fixed $m$ (e.g. $m = 2\times 10^{-17}$~eV for $\kappa = 0.1$), increasing the $\lambda$, eventually switching from attractive to repulsive interactions results in exiting the detectable region. This is because for repulsive interactions, the density of the system begins to drop, cf. Fig.~\ref{fig:Mfix}. For a large enough repulsive interaction strength (large positive $\lambda$), we thus expect that, at larger fixed $m$ (e.g. $m = 10^{-16}$~eV for $\kappa = 0.1$), an upper bound arises on the positive $\lambda$, from the fact that the density is too low to result in observable GW dephasing. Attractive interactions can therefore be probed more easily as a result, but up to a limit dictated by residing in the stable configuration regime. On the other hand, for fixed coupling strength, although it is favorable to increase the mass of DM $m$, as this corresponds to a higher density, the size of the boson star decreases quickly, as given in \eqref{eq:norm_fixM}, thereby not leaving enough inspiral region for the dephasing to accumulate. Increasing to $\kappa = 0.8$, the density of the system naturally increases, as seen in Sec.~\ref{subsec:fix_M}, opening up more regions for probing the DM mass. The detectable region of couplings shifts by two orders of magnitude due to the decrease in $M$ for larger $\kappa$, resulting in $a_M$ increasing, which can be discerned from \eqref{eq:norm_fixM}.

\section{Conclusion}
\label{sec:conclude}
We presented a comprehensive study of a system comprising a boson star hosting a black hole at its center, in the nonrelativistic limit. Our numerical analysis takes into careful consideration the gravitational potential of the central black hole, appropriately modifying the boundary conditions to obtain the solutions in hydrostatic equilibrium. The equilibrium configurations arise from a balance between the quantum pressure, the pressure arising from the self-gravity of the boson star and the black hole gravitational potential, and the pressure generated from the self-interactions of DM, if present. We find that for repulsive or no interactions $(\lambda \geq 0)$, all the equilibria are stable, corresponding to BS-BH systems that become less dense and larger in size as the strength of the coupling increases. In contrast, attractive interactions $(\lambda < 0)$ lead to unstable equilibria in regions of parameter space, setting constraints on the maximum mass of the boson star, for a given $|\lambda|$, or on the minimum coupling for fixed boson star mass, with these bounds being dependent on the mass of the central black hole. For the stable configurations, the boson star is denser and more compact. Overall, the presence of a black hole enhances the density of the boson star while shrinking it, with the effect being more pronounced as the mass of the black hole increases. These results, revealing the full parameter space, are presented in Figs.~\ref{fig:afix_att_rep} and \ref{fig:Mfix}, and are given in terms of normalized quantities, allowing one to easily extract the physical quantities of interest.

We then consider an approximate form of the density profile compatible with the boundary conditions of the exact numerical solution, the ansatz given in Eq.~\eqref{eq:ansatz}. Following an approach based on minimizing the effective potential of the system, we derive the mass-radius relation of the BS-BH system given in Eq.~\eqref{eq:mass-rad_rel}. Compared to our numerical results, we found good qualitative agreement, with quantitative agreement within a few percent for attractive interactions. For repulsive interactions, we find similar quantitative agreement when the ratio between the black hole and boson star masses $\kappa < 0.3$. This suggests applications of our approximate results in the context of DM halos around BHs, with a core comprising a self-gravitating Bose-Einstein condensate with attractive interactions, such as for axionlike particles.

Finally, we study the phenomenology of the system in the context of gravitational waves emitted from the extreme-mass ratio inspiral of this system with a second black hole. The presence of the DM environment results in additional energy loss during the merger, inducing GW dephasing. To compute the energy lost, we apply a formula for dynamical friction that accounts for the quantum pressure of the light scalar particles around the second black hole. Through a Fisher matrix forecast, we reveal the parameter space of the mass of ULDM and coupling that may be detected based on half a year of observation time by LISA in Fig.~\ref{fig:constraints}, based on the GW dephasing effect.

\section*{Acknowledgments}
A.B. and J.H.K. are supported partly by the National Research Foundation of Korea (NRF) Grant No. NRF-2021R1C1C1005076, RS-2026-25484206, the BK-21 FOUR program through NRF, and the Institute of Information \& Communications Technology Planning \& Evaluation (IITP)-Information Technology Research Center (ITRC) Grant No. IITP-2025-RS-2024-00437284. A.B. also receives support from the NRF, under Grant No. NRF-2020R1I1A3068803.
X.-Y.~Y. is supported in part by the KIAS Individual Grant No.~QP090702. We acknowledge the hospitality at APCTP where part of this work was done.

\appendix
\numberwithin{equation}{section}

\section{Comparison with the Relativistic Treatment}
\label{app:comparison}

In this Appendix, we present a comparison between the results of the nonrelativistic approach through GPP equations presented in the main text, and the relativistic treatment based on solving the Einstein-Klein-Gordon equations, as presented in Ref.~\cite{Barranco:2017aes}. We focus on the case with no self-interactions, as the impact of these is to result in sparser (repulsive interactions) or denser (attractive interactions) BS-BH systems, and therefore, we expect the trends, discussed in detail in Sec.~\ref{sec:density_profile}, to generalize accordingly.

\begin{figure}[htbp]
    \centering
    \includegraphics[width=0.48\textwidth]{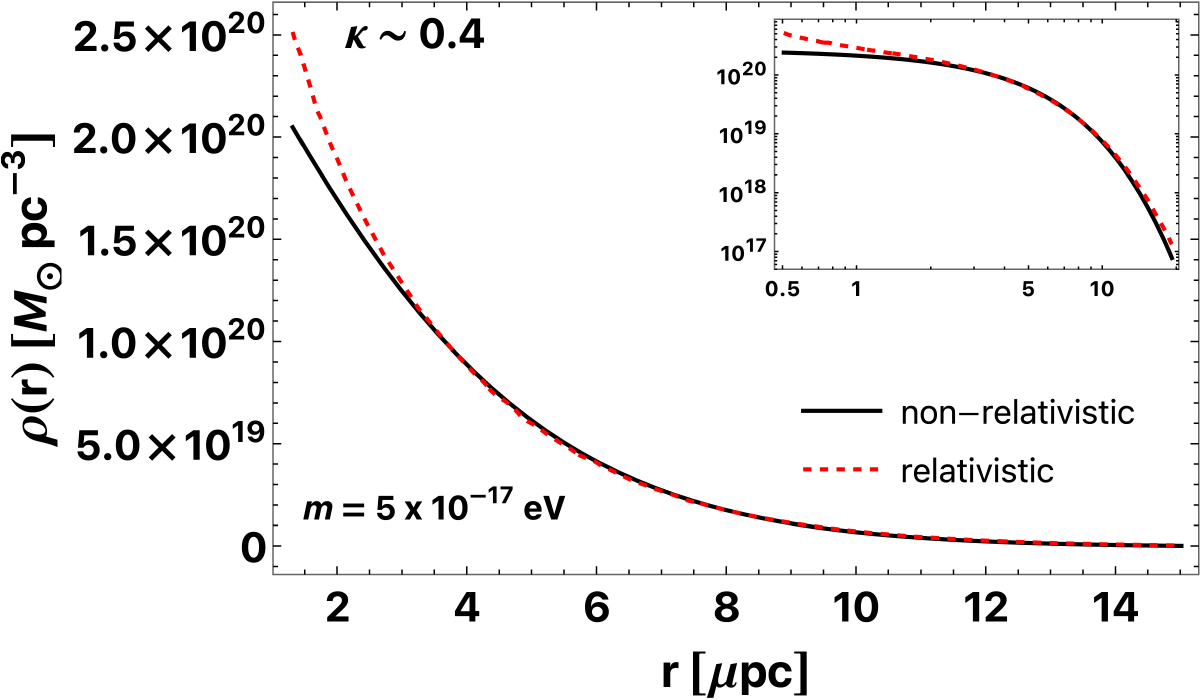}
    \caption{Comparison of density profiles for a BS-BH system without self-interactions obtained through the nonrelativistic treatment considered in this work (solid black lines) and with the full relativistic treatment (dashed red lines) of Ref.~\cite{Barranco:2017aes} for the benchmark case of $\kappa \sim 0.4$. \textit{Inset}: The same density profiles but in log-log scale. We have taken the scalar mass $m = 5\times10^{-17}$~eV, and the central black hole mass $ M_{\rm BH} = 10^5~M_{\odot}$. The full relativistic treatment indicates a larger central density, but the density profiles agree well (at a few percent level) for larger radii.
    }
    \label{fig:relcomparison}
\end{figure}

In Fig.~\ref{fig:relcomparison}, we compare the density profile obtained by solving Eq.~\eqref{eq:hydro} with the corresponding solution in Ref.~\cite{Barranco:2017aes}, which considers the set of relativistic equations, for $\kappa \sim 0.4$. The two density profiles agree well, within $\sim 5\%$, for larger radii, but deviate at smaller radii, indicating that relativistic effects can enhance the central density by factors of order one. Correspondingly, for the purpose of GW observations, as discussed in Sec.~\ref{sec:gw_probe}, we expect that this difference in the central density would not affect our main inferences on the parameter space, on account of the dephasing effect being more pronounced at lower frequencies, i.e. at larger radii.

\section{Formulae for the Ansatz}
\label{app:ansatz_analysis}
\begin{figure}[htbp]
    \centering
    \includegraphics[width=0.4\textwidth]{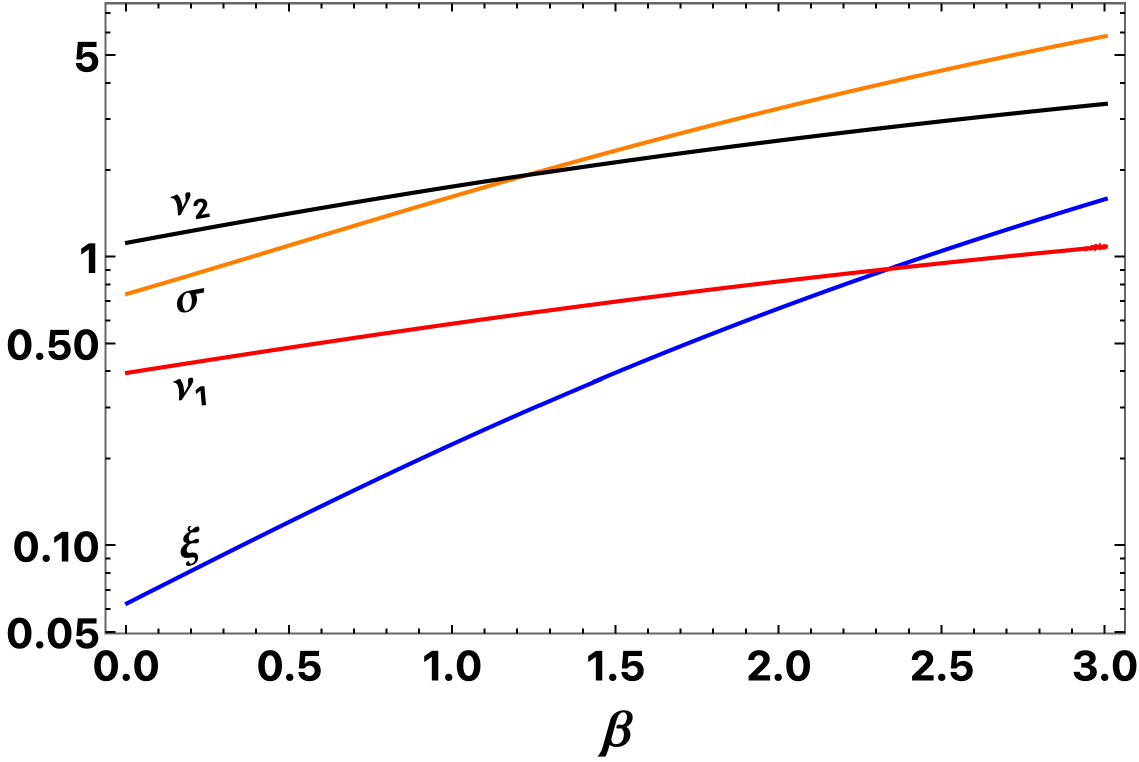}
    \caption{Various coefficients given in Eqs.~\eqref{eq:coeff} as functions of $\beta = G M_{\rm BH}m^2 R$. For fixed $R$ and $m$, increasing $\beta$ implies increasing the $M_{\rm BH}$. For $\beta = 0$, the coefficients give the values $\sigma(0) = 0.75$, $\xi(0) \approx 0.0635$, $\nu_1(0) \approx 0.399$ and $\nu_2(0) \approx 1.128$.
    }
    \label{fig:coeff}
\end{figure}

\begin{figure}[htbp]
    \centering
    \includegraphics[width=0.4\textwidth]{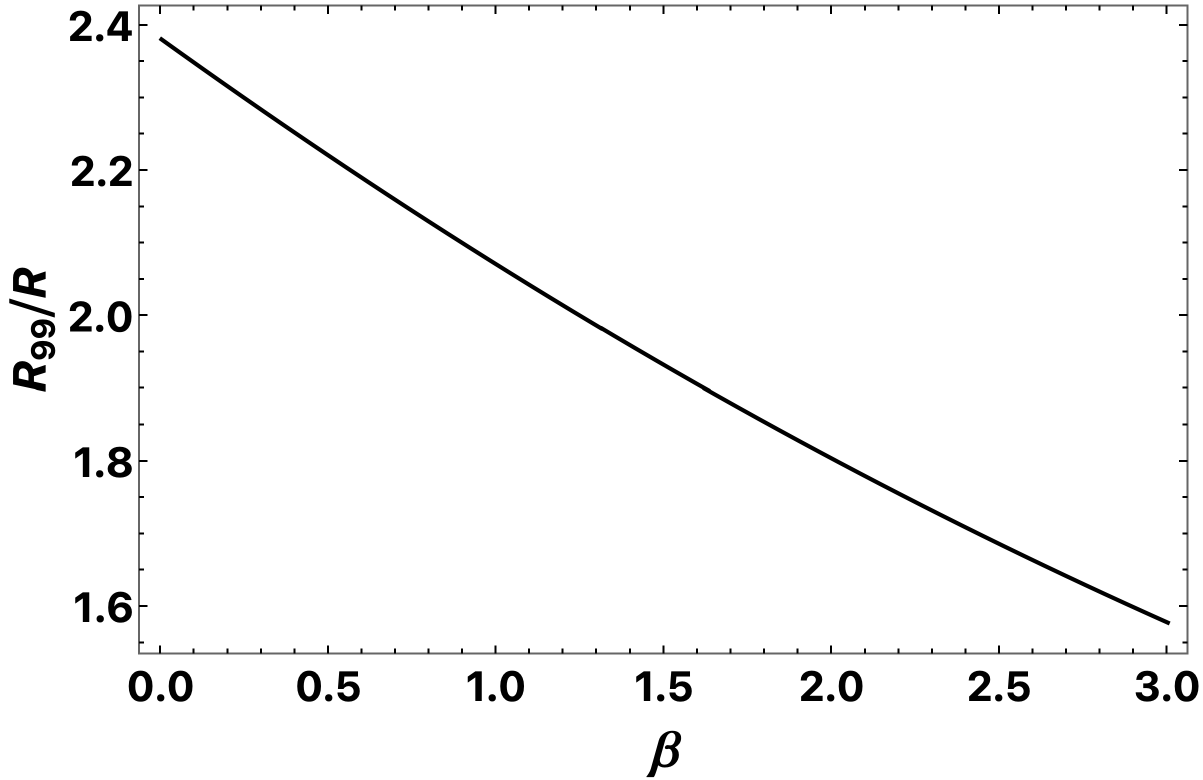}
    \caption{Radius containing 99\% of the boson star mass $M$ scaled to the characteristic radius $R$ as a function of $\beta$, based on the ansatz~\eqref{eq:ansatz}. For $\beta = 0$, we recover the usual value of $\sim 2.382$, i.e., a pure Gaussian ansatz, applicable when there is no central black hole within the boson star.
    }
    \label{fig:R99}
\end{figure}

In Sec.~\ref{sec:ansatz}, based on the ansatz for the density profile in Eq.~\eqref{eq:ansatz}, we computed an effective potential for the BS-BH system. In this Appendix, we provide expressions for the coefficients pertaining to the quantum pressure~\eqref{eq:quantum_pot}, the internal energy~\eqref{eq:internal_energy}, and the potential for the gravitating BS~\eqref{eq:grav_energy_BS} and the BS-BH gravitational interaction~\eqref{eq:grav_energy_BH}. We first define
\begin{equation}
    D(\beta)\equiv -2\beta + e^{\beta^2}\sqrt{\pi}(1+2\beta^2)\,{\rm erfc}(\beta)\,,
\end{equation}
where ${\rm erfc}(x) \equiv 1 - {\rm erf}(x)$, with ${\rm erf}(x)$ being the error function. This then gives,
\begin{subequations}
    \begin{equation}
        \sigma(\beta) = \frac{1}{4}\,\left[1+\frac{2\sqrt{\pi}\,\beta^2 \,e^{\beta^2}\,{\rm erfc}(\beta)}{D(\beta)}\right]\,,
    \end{equation}
    \begin{equation}
        \xi(\beta) = \frac{1}{4}\,\left[\frac{-4\beta + e^{2\beta^2}\sqrt{\pi}(1+4\beta^2)\,{\rm erfc}(\sqrt{2}\beta)}{D^2(\beta)}\right]\,,
    \end{equation}
    \begin{equation}
        \begin{aligned}
            \nu_1(\beta) &=
            \frac{\sqrt{\pi}\,e^{\beta^2}}{2D^2(\beta)} \bigg\{\beta\big[
                    -2 - 4 \beta^2
                    + 2 (1 + 2\beta^2) \, {\rm erf}(\beta)^2 \\
                         &\quad + 4 (1 + 2\beta^2) \, {\rm erfc}(\beta)
                         - 4 (1 + 2\beta^2) \, {\rm erfc}(\beta)^2
                     \big] \\
                         &\quad + 8 \beta^2 - 8 \beta^2 \, {\rm erf}(\beta) \\
                         &\quad + \sqrt{2} e^{\beta^2} \left[
                     {\rm erfc}(\sqrt{2} \beta) - 4 \beta^2 {\rm erfc}(\sqrt{2} \beta)\right] \bigg\}\,,
        \end{aligned}
    \end{equation}
    \begin{equation}
        \nu_2(\beta) = \frac{2(-1 + e^{\beta^2}\sqrt{\pi}\,{\rm erfc}(\sqrt{2}\beta))}{D^2(\beta)}\,.
    \end{equation}
    \label{eq:coeff}
\end{subequations}
These coefficients are plotted as functions of $\beta$ in Fig.~\ref{fig:coeff}. Additionally, we give the coefficient specifying the radius containing $99\%$ of the boson star mass in Fig.~\ref{fig:R99}.

\section{LISA Detector Response Functions}
\label{app:LISA_detect}
We provide here the detector response functions used to calculate the GW strain in LISA as given in \eqref{eq:GWstrain_det}. We have
\begin{subequations}
    \begin{align}
        F_+ & \equiv \frac12\left[D_+\cos2\varsigma-D_\times\sin2\varsigma\right]\,,
        \\
        F_\times &\equiv \frac12\left[D_+\sin2\varsigma+D_\times\cos2\varsigma\right]\,,
    \end{align}
\end{subequations}
with
\begin{subequations}
    \begin{equation}
        \begin{aligned}
            D_{+}
            &= \frac{\sqrt{3}}{64} \Big\{
                -36 \sin^2\vartheta\, \sin(2\alpha - 2\beta) \\
            &\quad - 4\sqrt{3} \sin 2\vartheta \big[\sin(3\alpha - 2\beta - \varphi) - 3\sin(\alpha - 2\beta + \varphi)\big] \\
            &\quad + [\cos 2\vartheta + 3]\big[\cos 2\varphi (9\sin 2\beta - \sin(4\alpha - 2\beta)) \\
            &\quad+ \sin 2\varphi (\cos(4\alpha - 2\beta) - 9\cos 2\beta)\big] \Big\},
        \end{aligned}
    \end{equation}
    \begin{equation}
        \begin{aligned}
            D_{\times}
                &= \frac{1}{16} \Big\{
                    \sqrt{3}\cos\vartheta\,[9\cos(2\beta - 2\varphi) - \cos(4\alpha - 2\beta - 2\varphi)] \\
                &\quad - 6\sin\vartheta\,[\cos(3\alpha - 2\beta - \varphi) + 3\cos(\alpha - 2\beta + \varphi)]
            \Big\}.
        \end{aligned}\label{eq:Dcross}
    \end{equation}
\end{subequations}
Here, $\alpha = 2\pi t + \alpha_0$ is the orbital phase of the guiding center, and $\beta = 2\pi n/3 + \beta_0$, with $n = 0,1,2$ for three spacecrafts, is the relative phase of the spacecraft within the constellation. The parameters $\alpha_0$ and $\beta_0$ give the initial ecliptic longitude and orientation of the constellation. Finally, the delay between the arrival time of GWs at the Sun and the arrival time at the detector $\Delta t$, given by
\begin{equation}
    \Delta t = -\frac{1\,{\rm AU}}{c}\,\sin\vartheta\cos(\alpha-\varphi)\,,
\end{equation}
where ${\rm AU}$ refers to Astronomical Units.

\vspace{3\baselineskip}

\bibliography{references}

\end{document}